%% file: surf_op_Z2.tex
\pdfoutput=1
\documentclass[11pt,a4paper]{article}
\usepackage[accsupp]{axessibility}
\usepackage{cite}
\usepackage{graphicx}
\usepackage{amssymb}
\usepackage{amsmath}
\usepackage{amsfonts}
\usepackage{dsfont}
\usepackage{mathtools}
\usepackage{slashed}
\usepackage{rotating}
\usepackage{bbold,amsfonts}

\usepackage[utf8]{inputenc}
\usepackage{bm}
\usepackage{xcolor}
\usepackage{float}
\usepackage{braket}

\usepackage[height=8.8in,width=6.45in]{geometry}
\usepackage[font=small,labelfont=bf]{caption}
\usepackage[hidelinks]{hyperref}
\numberwithin{equation}{section}

\newcommand{\beq}{\begin{equation}}
\newcommand{\eeq}{\end{equation}}

\DeclareMathOperator{\Tr}{Tr}

\newcommand{\ii}{\mathrm{i}}
\newcommand{\overbar}[1]{\mkern 1.5mu\overline{\mkern-1.5mu#1\mkern-1.5mu}\mkern 1.5mu}
\makeatletter
\newcommand*{\letterdef@}{}
\newcommand*{\letterdef}[3]{%
	\def\letterdef@##1{\expandafter\newcommand\csname #1\endcsname{#2{##1}}}%
	\@tfor\@tempa :=#3\do{\expandafter\letterdef@\expandafter{\@tempa}}}
\makeatother
\letterdef{c#1} {\mathcal}{ABCDEFGHIJKLMNOPQRSTUVWXYZ} 
\letterdef{rm#1}{\mathrm} {dDeimM} 

\begin{document}
\begin{titlepage}
\vbox{
    \halign{#\hfil         \cr
           } 
      }  
\vspace*{15mm}
\begin{center}
{\LARGE \bf 
Surface Defects from Fractional Branes - I
}

\vspace*{15mm}

{\Large S.~K.~Ashok$\,{}^{a}$,  M.~Bill\`o$\,{}^{b,c}$, M.~Frau$\,{}^{b,c}$,
	A.~Lerda$\,{}^{d,c}$ and S.~Mahato$\,{}^{a}$}
\vspace*{8mm}

${}^a$ Institute of Mathematical Sciences, \\
          Homi Bhabha National Institute (HBNI),\\
		 IV Cross Road, C.I.T. Campus, \\
		  Taramani, Chennai, India 600113
		  \vskip 0.5cm
			
${}^b$ Universit\`a di Torino, Dipartimento di Fisica,\\
		   Via P. Giuria 1, I-10125 Torino, Italy
		   \vskip 0.5cm

${}^c$ I.N.F.N. - sezione di Torino,\\
		   Via P. Giuria 1, I-10125 Torino, Italy 
		   \vskip 0.5cm		

${}^d$  Universit\`a del Piemonte Orientale,\\
			Dipartimento di Scienze e Innovazione Tecnologica\\
			Viale T. Michel 11, I-15121 Alessandria, Italy			
	
\vskip 0.8cm
	{\small
		E-mail:
		\texttt{sashok,sujoymahato@imsc.res.in; billo,frau,lerda@to.infn.it}
	}
\vspace*{0.8cm}
\end{center}

\begin{abstract}

We show that the Gukov-Witten monodromy defects of supersymmetric Yang-Mills 
theory can be realized in perturbative string theory by considering an orbifold 
background of the Kanno-Tachikawa type and placing stacks of fractional D3-branes 
whose world-volume partially extends along the orbifold directions. In particular, we 
show that turning on a constant background value for some scalar fields in the closed 
string twisted sectors induces a non-trivial profile for the gauge field and one of the 
complex scalars of the world-volume theory, and that this profile exactly matches the 
singular behavior that one expects for a Gukov-Witten surface defect in the 
${\mathcal N}=4$ super Yang-Mills theory. To keep the presentation as simple as 
possible, in this work we restrict our analysis to surface defects corresponding to a 
$\mathbb{Z}_2$ orbifold and defer the study of the most general case to a
companion paper.

\end{abstract}
\end{titlepage}

\tableofcontents
\vspace{1cm}
\begingroup
\allowdisplaybreaks

\section{Introduction}
\label{secn:intro}

Non-local operators have traditionally played an important role in quantum field theory
since they can provide valuable information especially at the non-perturbative level.
In this work we consider surface defects in 4$d$ 
supersymmetric Yang-Mills theories and study their explicit realization in string theory. 

From the gauge theory point of view, there are several ways to analyze surface defects. 
The original approach of Gukov and Witten (GW) in \cite{Gukov:2006jk,Gukov:2008sn} 
was to treat them as monodromy defects, in which one specifies the singular 
behavior of the fields of the gauge theory as one 
approaches the defect. Another possibility is to describe a surface defect 
as a 2$d$ quiver gauge theory with some degrees of freedom
coupled to a 4$d$ gauge theory \cite{Gaiotto:2009fs, Gaiotto:2013sma}. 
In many cases these two different descriptions lead to the same results \cite{Jeong:2018qpc}. For example,
the low-energy effective action on the Coulomb branch of the 4$d$ gauge theory 
computed in the two approaches exactly match. Moreover, by fruitfully combining the 
two methods various properties of the surface defects 
as well as many duality relations and non-perturbative effects can be studied 
\cite{Alday:2010vg,Gomis:2014eya,Frenkel:2015rda,Pan:2016fbl,Ashok:2017odt,
Gorsky:2017hro,Ashok:2017lko,Ashok:2018zxp,Ashok:2019rwa}. 

There are also several ways to embed the surface defects in string theory and, more generally, to study the defects from a higher-dimensional perspective. In \cite{Gomis:2007fi,Drukker:2008wr}
the GW defects were given a holographic representation in terms of 
bubbling geometries, which are particular solutions of Type II B supergravity that asymptote to 
$AdS_5\times S_5$.  
Since many 4$d$ supersymmetric gauge theories can be obtained by compactification 
from the 6$d$ $(2,0)$ theory defined on the world-volume of an M5 brane 
\cite{Witten:1997sc, Gaiotto:2009we}, surface defects can also be realized 
by introducing intersecting M5 branes or an M2 brane inside the M5 brane \cite{Frenkel:2015rda}.   
From this six dimensional perspective, surface defects have been recently studied in detail 
\cite{Balasubramanian:2018pbp} following earlier work in \cite{Gaiotto:2009we,
Gaiotto:2009hg, Chacaltana:2012zy}, by exploiting the relation to the Hitchin 
integrable system, with the aim of obtaining a complete classification of the surface 
defects in the 6$d$ theory. 
 
In this paper we shall study the simplest case of GW defects in the maximally 
supersymmetric ${\mathcal N}=4$ Yang-Mills theory with gauge group U(N) or SU(N). 
Our primary goal is to realize these surface defects in perturbative string theory and to 
recover the singular profiles of the fields in the gauge theory. We do so by calculating 
perturbative open/closed string amplitudes in Type II B string theory on an orbifold 
background.
Following a proposal of Kanno and Tachikawa (KT) \cite{Kanno:2011fw}, 
we engineer the 4$d$ ${\mathcal N}=4$ Yang-Mills theory by means of fractional D3-branes 
with two world-volume directions along the orbifold, leaving unbroken 
the Poincar\'e symmetry in the other two world-volume directions.  
This is quite different from the more familiar configuration 
in which the fractional D3-branes are completely transverse to the orbifold 
\cite{Douglas:1996sw}. In fact, in this latter case the resulting gauge theory has
Poincar\'e symmetry in four dimensions but a reduced amount of supersymmetry
since only a fraction of the sixteen supercharges of the orbifold background is 
preserved on the world-volume.

This orbifold set-up has already been studied in earlier works on the subject 
\cite{Alday:2010vg, Awata:2010bz, Kanno:2011fw, Ashok:2017odt} where also fractional 
D(--1)-branes have been introduced on top of the fractional D3-branes to derive the so-called 
ramified instanton partition function in the presence of a surface operator, extending the equivariant localization methods of \cite{Nekrasov:2002qd}.
In this paper, instead, we consider only stacks of fractional D3-branes and focus on the gauge theory 
defined on their world-volume, which has largely remained unexplored. In particular
we compute correlators involving both the massless fields of the gauge theory and 
the massless twisted scalars in the Neveu-Schwarz/Neveu-Schwarz (NS/NS) and 
Ramond/Ramond (R/R) sectors of the closed string background, and show that these correlators
precisely encode the singular profiles that define a GW defect. The continuous parameters that appear in these profiles and that are part of the defining data of a surface operator are related to the vacuum expectation values of the twisted scalars. In this way we clarify in detail how the KT set-up realizes surface defects in the gauge theory. 

We believe that this construction of surface defects using perturbative string theory is interesting because it provides an explicit and calculable framework which may be useful also for various generalizations and applications. For instance, by introducing orientifolds, it would be possible to generalize our results to other classical gauge groups. Similarly, one could try to construct surface defects in quiver gauge theories or in theories with less supersymmetry. Furthermore, our explicit string realization of surface defects may turn out to
be useful in computing other quantities that characterize the superconformal defect field theory
and help to establish connections with alternative approaches to the study of defects.

This paper is organized as follows: in Section~\ref{sec:GWproposal}, we review the main features of monodromy defects and outline our proposal to recover the properties of the defect using fractional branes. In Section~\ref{KTorbifold} we describe in more detail the orbifold background and discuss the massless fields of the twisted sectors that will play an important role in our analysis. To avoid too many technical issues, we consider only the case of simple defects deferring the analysis of 
the general case to a companion paper \cite{Ashok:2020jgb}. In Section~\ref{sec:fD3} we introduce the fractional branes that realize the surface defects. In 
Section~\ref{correlators} we go on to compute the relevant disk correlators and then in 
Section~\ref{sec:profile} we show how these can be used to derive the singular profiles of the gauge fields and scalars in the presence of the defect. This allows us to relate the continuous parameters of the surface defects to expectation values of certain twisted closed string fields.
Finally in the concluding section we comment on how our results are consistent with the action of S-duality in the ${\mathcal N}=4$ gauge theory. We collect some technical material in the Appendices.

\section{Review and outline}
\label{sec:GWproposal}
We begin by briefly reviewing the main features of the GW monodromy defects 
in ${\mathcal N}=4$ super Yang-Mills theory with gauge group U$(N)$ or SU($N$), 
following \cite{Gukov:2006jk,Gukov:2008sn}\,%
\footnote{See also the review \cite{Gukov:2014gja}.}. 
Then we outline the main ideas behind our proposal to complete 
the KT description \cite{Kanno:2011fw} of such defects in terms of fractional D3-branes 
in an orbifold background. The details of our proposal will be fully discussed in the 
following sections.

\subsection{Monodromy defects}
\label{subsec:md}
Let us consider an $\cN=4$ super Yang-Mills theory defined on 
$\mathbb{R}^4 \simeq \mathbb{C}_{(1)}\times\mathbb{C}_{(2)}$. We will use 
the complex coordinate $z_i$ on $\mathbb{C}_{(i)}$; we will also use polar coordinates 
in $\mathbb{C}_{(2)}$ setting $z_2 = r\,\rme^{\ii\theta}$. It will also be useful at times 
to denote by $\vec x_\parallel$ and $\vec x_\perp$ the real coordinates of these two 
planes and by $\vec k_\parallel$ and $\vec k_\perp$ the corresponding momenta.

A monodromy defect $D$ is a surface defect extended along $\mathbb{C}_{(1)}$ and 
placed at the origin of $\mathbb{C}_{(2)}$. It is defined by the singular behavior of some 
of the bosonic fields of the theory, namely the 1-form gauge connection $\mathbf{A}$ 
and one of the three complex adjoint scalars, which we call $\mathbf{\Phi}$. 
Near the location of the defect, {\it{i.e.}} for $r\to 0$, these fields have the following 
non-trivial profile:
\begin{equation}
\mathbf{A} = \begin{pmatrix}
		\alpha_0\,\mathbb{1}_{n_0}&0&\cdots&0\\
		0&\alpha_1\,\mathbb{1}_{n_1}&\cdots&0\\
		\vdots&\vdots&\ddots&\vdots\\
		0&0&\cdots&\alpha_{M-1}\,\mathbb{1}_{n_{M-1}}
		\end{pmatrix} d\theta ~,
		\label{Aprofile}
\end{equation}
and
\begin{equation}
\mathbf{\Phi} = \begin{pmatrix}
		(\beta_0+\ii\,\gamma_0)\,\mathbb{1}_{n_0}&0&\cdots&0\\
		0&(\beta_1+\ii\,\gamma_1)\,\mathbb{1}_{n_1}&\cdots&0\\
		\vdots&\vdots&\ddots&\vdots\\
		0&0&\cdots&(\beta_{M-1}+\ii\,\gamma_{M-1})\,\mathbb{1}_{n_{M-1}}
		\end{pmatrix} \frac{1}{2z_2} ~.
		\label{Phiprofile}
\end{equation}
Here $\mathbb{1}_{n_I}$ denotes the $(n_I\times n_I)$ identity matrix; 
$\alpha_I$, $\beta_I$ and $\gamma_I$ are real parameters and the integers $n_I$ are 
such that
\begin{equation}
\sum_{I=0}^{M-1} n_I = N~.
\label{partition}
\end{equation}
This non-trivial field configuration breaks the U$(N)$ gauge group to a Levi subgroup
\begin{align}
	\label{levi}
		\mathrm{U}(n_0)\times \mathrm{U}(n_1)\times \cdots 
		\times \mathrm{U}(n_{M-1})~.
\end{align}
If the gauge group is SU($N$) one has to remove the overall U(1) factor from (\ref{levi})
and subtract the trace from (\ref{Aprofile}) and (\ref{Phiprofile}).

In the definition of the path-integral one is allowed to turn on a 2$d$ $\theta$-term, 
whose coefficient we denote $\eta_I$ for each factor in the unbroken Levi subgroup. This 
means that in the path-integral we include the following phase factor: 
\begin{align}
	\label{thetaterm}
		\exp\left(\ii\,\sum_{I=0}^{M-1} \eta_I \int_D  \Tr_{\mathrm{U}(n_I)} F_I\right)~.
\end{align}
Altogether, we can say that a monodromy defect is characterized by the 
discrete parameters $n_I$, which constitute a partition of $N$, and by the four sets 
of real continuous parameters 
$\{\alpha_I,\beta_I,\gamma_I,\eta_I\}$, with $I=0,\ldots,M-1$.  

One of the remarkable features of the $\mathcal{N}=4$ Yang-Mills theory is its 
invariance under the action of the non-perturbative duality group SL$(2,\mathbb{Z})$. It 
turns out that this duality also acts naturally on the parameters of the surface operator as 
shown in \cite{Gukov:2006jk}. In particular, an element 
$\Lambda=\bigl(\begin{smallmatrix}
m & n\\
p & q\\
\end{smallmatrix}\bigr) \in \text{SL}(2,\mathbb{Z})$ induces the transformation
\begin{equation}
\begin{aligned}
(\alpha_I,\eta_I)&~\longrightarrow ~ (\alpha_I,\eta_I) \,\Lambda^{-1} 
= (q\,\alpha_I-p\,\eta_I,
-n\,\alpha_I+m\,\eta_I)~,\\
(\beta_I,\gamma_I)&~\longrightarrow ~|p\,\tau+q|\,(\beta_I,\gamma_I)
\end{aligned}
\label{Sdualityonabge}
\end{equation}
where $\tau$ is the complexified gauge coupling constant.

\subsection{Monodromy defects from fractional branes}
\label{proposal}
Our primary goal is to derive the field profiles (\ref{Aprofile}) and
(\ref{Phiprofile}) as well as the topological term (\ref{thetaterm}) that characterize a 
monodromy GW defect from a world-sheet analysis of its orbifold realization proposed 
in \cite{Kanno:2011fw}. 

In this set-up, the gauge theory lives on a system of D3-branes in Type II B string theory
placed in a $\mathbb{Z}_M$ orbifold space. The orbifold group acts on two complex 
planes $\mathbb{C}_{(2)}\times \mathbb{C}_{(3)}$, the first of which is transverse to the 
defect inside the world-volume of the D3-branes, while the second is transverse to 
the D3's. In this realization, therefore, the defect $D$ is located at the fixed point of the 
orbifold action. The integer partition of $N$, (see (\ref{partition})), which 
determines the unbroken  Levi subgroup (\ref{levi}) corresponds to the choice of 
the $N$-dimensional representation of $\mathbb{Z}_M$ on the Chan-Paton indices 
of the D3-branes; in other words, $n_I$ is the number of the fractional branes 
transforming in the $I$-th irreducible representation of $\mathbb{Z}_M$. We shall refer to these fractional branes as  
D3-branes of type $I$.

What is missing in the KT description is how the orbifold realization encodes the 
continuous parameters of the monodromy defect. Our goal is to fill this gap by showing 
that they correspond to background values for fields belonging to the twisted sectors of 
the closed string theory on the orbifold. In particular, the twisted background fields in 
the NS/NS sector, which here we collectively denote as $b$, account for the parameters 
$\alpha_I$, $\beta_I$ and $\gamma_I$ which appear in (\ref{Aprofile}) and 
(\ref{Phiprofile}), while the twisted scalar of the R/R sector, which we denote $c$, accounts 
for the parameters $\eta_I$ in the topological term (\ref{thetaterm}).

Schematically, the mechanism goes as follows. In the presence of a closed string background 
certain open string fields $\Phi_{\mathrm{open}}$ attached to a fractional D3-brane of 
type $I$ acquire a non-zero one-point function, {\it{i.e.}} a tadpole. 
If we denote by $\cV_{\mathrm{open}}$ the open string vertex operator associated to 
$\Phi_{\mathrm{open}}$ and by $\mathcal{V}_b$ the closed string vertex operator
corresponding to $b$, the tadpole 
$\big\langle \mathcal{V}_{\text{open}} \big\rangle_{b;I}$ 
arises from an open/closed string correlator evaluated on a disk which contains
an insertion of $b\,\mathcal{V}_b$ in the interior and of the vertex operator
$\cV_{\mathrm{open}}$ on the boundary that lies on a D3-brane of type $I$:
\begin{align}
	\label{tadpole}
	\big\langle \mathcal{V}_{\text{open}} \big\rangle_{b;I}\,\equiv~~~
		\parbox[c]{.29\textwidth}{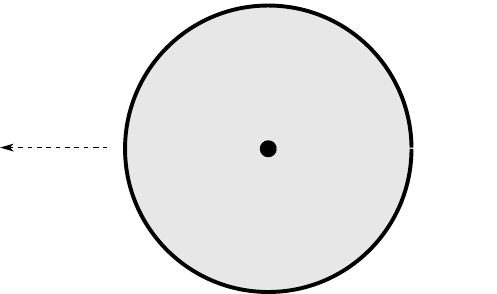}~.
\end{align} 
Note that the open string vertex carries momentum along the D3-brane world-volume. While 
its longitudinal components $\vec k_\parallel$ along the defect are set to zero by momentum conservation, its transverse components $\vec k_\perp$ need not be set to zero, 
as we have pictorially indicated in the diagram. 
Indeed, the twisted fields, which are localized at the 
orbifold fixed point, break translation invariance along the orbifold directions and thus 
$\vec k_\perp$ does not need to be conserved. Therefore, the disk diagram represented 
above acts as a classical source for $\Phi_{\mathrm{open}}$, which acquires a 
non-trivial profile in the plane transverse to the defect. The explicit expression of this 
profile near the defect is obtained by attaching a propagator to the source 
and taking the Fourier transform ($\mathcal{FT}$), namely
\begin{align}
	\label{Born}
		\Phi_{\mathrm{open}}(\vec x_\perp) = 
		\mathcal{FT} \Big[\,
		\frac{1}{\vec k_\perp^2}\,\big\langle \cV_{\mathrm{open}}\big\rangle_{b;I}(\vec k_\perp)
		\,\Big]~.
\end{align}
The fields $\Phi_{\mathrm{open}}$ which get a tadpole from this mechanism arise 
from open strings with both ends on the same D3-brane, so they have diagonal 
Chan-Paton factors. In the following we will show in detail that the only 
non-zero tadpoles are those of the diagonal entries of the transverse components%
\footnote{We use the complex notation 
$\mathbf{A} = \sum_{i=1}^2\left(\mathbf{A}_i d\bar{z}_i+ \bar{\mathbf{A}}_i dz_i 
\right)$ to facilitate the comparison with (\ref{Aprofile}).} 
$\mathbf{A}_2$ and $\bar{\mathbf{A}}_2$ of the gauge connection 1-form and the 
diagonal entries of the complex scalar $\mathbf{\Phi}$. These are precisely the fields 
which have a non-trivial profile in a monodromy defect of GW type. Moreover, we 
will show that the functional dependence on the transverse coordinates acquired by 
these fields through (\ref{Born}) coincides with that of (\ref{Aprofile}) and 
(\ref{Phiprofile}), thereby identifying the parameters $\alpha_I$, $\beta_I$ and 
$\gamma_I$ with some of the background fields of the twisted NS/NS sector.   

The mechanism that encodes the non-trivial profile of the surface defect in a 
perturbative disk diagram is reminiscent of the way in which disks
with mixed D3/D(--1) boundary conditions account for the classical profile of the 
instanton solutions \cite{Billo:2002hm}. In that case, however, the defect is point-like 
and the classical profile of the fields depends on all world-volume coordinates; 
moreover, the role of parameters that appear in the profile is played by the D3/D(--1) open string moduli, instead 
of the closed string moduli, as in the present situation.

The twisted fields in the R/R sector also couple to the open string 
excitations through disk diagrams analogous to the one in (\ref{tadpole}). 
It turns out that the only non-zero diagrams of this type involve the 
diagonal entries of the longitudinal components $\mathbf{A}_1$ and 
$\bar{\mathbf{A}}_1$ of the gauge connection and do not depend on the transverse 
momentum $\vec k_\perp$. Thus, these diagrams are not tadpoles and do not lead 
to the emission of open string fields with a non-trivial profile; instead, they account for
some terms of the defect effective action, and in particular correspond 
to the $\theta$-terms of (\ref{thetaterm}). This implies that the parameters $\eta_I$ arise from 
the twisted R/R background fields.  

The description of the monodromy defects that we propose is analogous to the 
holographic description of defects given by \cite{Gomis:2007fi,Drukker:2008wr} in terms of 
bubbling geometries. Also in that case one gives a bulk description 
of the defect that accounts for all of its parameters in terms a closed string background. 
The orbifold description that we will discuss in the  following is however quite different since 
it makes use of perturbative string theory and world-sheet conformal field theory tools.

In the rest of the paper we will consider the particular case $M=2$, corresponding to 
simple defects. This case allows us to illustrate all the ingredients and mechanisms 
involved in our proposal, while at the same time avoiding some of the more technical issues related to the general ${\mathbb Z}_M$ orbifold. Therefore, in order to make the presentation more transparent, we shall discuss the generic surface defect in a companion paper \cite{Ashok:2020jgb}.

\section{Closed strings in the $\mathbb{Z}_2$ orbifold}
\label{KTorbifold}

We consider Type II B string theory propagating in a 10$d$ target space given by the 
orbifold
\begin{equation}
\mathbb{C}_{(1)}\times\frac{\mathbb{C}_{(2)}\times \mathbb{C}_{(3)}}{\mathbb{Z}_2}
\times \mathbb{C}_{(4)} \times \mathbb{C}_{(5)}~.
\label{orbifold}
\end{equation}
The $i$-th complex plane $\mathbb{C}_{(i)}$ is parametrized by the complex 
coordinates $z_i$ and $\bar{z}_i$ defined as
\begin{equation}
z_i=\frac{x_{2i-1} + \ii \,x_{2i}}{\sqrt{2}}\qquad\mbox{and}\qquad
\bar{z}_{i}=\frac{x_{2i-1} - \ii \,x_{2i}}{\sqrt{2}}
\label{coords}
\end{equation}
in terms of the ten real coordinates $x_\mu$, and the non-trivial element of the
$\mathbb{Z}_2$ orbifold group acts as follows
\begin{equation}
\label{orbact}
\begin{aligned}
(z_2, z_3) ~\longrightarrow~ (-z_2,  -z_3)\qquad\mbox{and}\qquad
(\bar{z}_{2}, \bar{z}_{3}) ~\longrightarrow~ (-\bar{z}_{2},  -\bar{z}_{3})~.
\end{aligned}
\end{equation}
This breaks the $\mathrm{SO}(4)\simeq\mathrm{SU}(2)_+\otimes \mathrm{SU}(2)_-$
isometry of the space $\mathbb{C}_{(2)}\times \mathbb{C}_{(3)}$ down to $\mathrm{SU}(2)_+$.

To describe closed strings in the orbifold (\ref{orbifold}) we use a complex notation 
analogous to the one in (\ref{coords}). We denote the left-moving bosonic 
string coordinates as $Z^i(z)$ and $\bar{Z}^i(z)$, and the right-moving ones as 
$\widetilde{Z}^i(\bar{z})$ and $\widetilde{\overbar{Z}^i}(\bar{z})$.
Here, $z$ and $\bar{z}$ are the complex coordinates that parametrize the world-sheet 
of the closed strings. In a similar way, we introduce the complex world-sheet fermionic 
coordinates $\Psi^i(z)$ and $\overbar{\Psi}^i(z)$, and their right-moving counterparts
$\widetilde{\Psi}^i(\bar{z})$ and $\widetilde{\overbar{\Psi}^i}(\bar{z})$. In all of our string computations we will use the convention that $2\pi\alpha^\prime=1$.

For the $\mathbb{Z}_2$ orbifold under consideration, the Hilbert space of the closed 
string, in addition to the usual untwisted sector, possesses  one twisted sector, associated 
to the non-trivial conjugacy class of $\mathbb{Z}_2$. In the following, we are going to 
briefly review%
\footnote{See for instance \cite{Anselmi:1993sm,Douglas:1996sw,Billo:2001vg} for more detailed accounts of various properties of the CFT on a $\mathbb{C}^2/\Gamma$ orbifold space.} the main properties of this twisted sector which will play a crucial role in 
our analysis.

\subsection{Twisted closed string sectors}
\label{subsec:tcc}
In the twisted sector the left-moving bosonic string coordinates $Z^2$ and $Z^3$ 
are anti-periodic:
\begin{align}
	\label{twistcond}
		Z^2(\rme^{2\pi\ii}\,z) & = -\, Z^2(z)\qquad\mbox{and}\qquad
		Z^3(\rme^{2\pi\ii}\,z) =  -\, Z^3(z) ~.
\end{align}
Of course, the same happens for the complex conjugate coordinates 
$\overbar{Z}^2(z)$ and $\overbar{Z}^3(z)$.
The vacuum for these twisted bosonic fields is created by the operator 
\begin{equation}
\Delta(z)=\sigma^2(z)\,\sigma^3(z)~,
\label{Delta}
\end{equation}
where $\sigma^2(z)$ and $\sigma^3(z)$ are the twist fields \cite{Dixon:1986qv} in the 
complex directions 2 and 3. Each of these twist fields is a conformal operator of weight 
1/8 so that $\Delta(z)$ has weight 1/4.

A completely analogous construction can be made in the right-moving sector, where one 
has
\begin{align}
	\label{twistcondr}
		\widetilde{Z}^2(\rme^{2\pi\ii}\,\bar{z}) & = -\,
		\widetilde{Z}^2(\bar{z})\qquad\mbox{and}\qquad
		\widetilde{Z}^3(\rme^{2\pi\ii}\,\bar{z}) =  -\,\widetilde{Z}^3(\bar{z})~,
\end{align}
and similarly for their complex conjugates. Correspondingly, one defines the 
right-moving twist field $\widetilde{\Delta}(\bar{z})$ of dimension 1/4.

As far as the fermionic coordinates are concerned, one has
\begin{align}
	\label{twistcondf}
		\Psi^2(\rme^{2\pi\ii}\,z) = \mp \, \Psi^2(z)\qquad\mbox{and}\qquad
		\Psi^3(\rme^{2\pi\ii}\,z) = \mp \, \Psi^3(z)~,		
\end{align}
where the upper signs refer to the NS sector and the lower ones to the R sector. 
The complex conjugate coordinates $\overbar{\Psi}^2(z)$ and $\overbar{\Psi}^3(z)$
have similar monodromy properties.
In the right-moving sector, the fermionic fields are such that
\begin{align}
	\label{twistcondfrm}
		\widetilde{\Psi}^2(\rme^{2\pi\ii}\bar{z}) = \mp \,\widetilde{\Psi}^2(\bar{z})
		\qquad\mbox{and}\qquad
		\widetilde{\Psi}^3(\rme^{2\pi\ii}\bar{z}) 
		= \mp \,\widetilde{\Psi}^3(\bar{z})~,		
\end{align}
with similar expressions for the complex conjugate coordinates 
$\widetilde{\overbar{\Psi}^2}(\bar{z})$ and $\widetilde{\overbar{\Psi}^3}(\bar{z})$.

As a consequence of these monodromy properties, in the expansion of the various 
fields the moding is shifted by $1/2$ with respect to their untwisted values. In particular, 
the bosonic fields $Z^2$, $Z^3$, $\overbar{Z}^2$ and $\overbar{Z}^3$ have half-integer modes, 
while the fermions $\Psi^2$, $\Psi^3$, $\overbar{\Psi}^2$ and $\overbar{\Psi}^3$ have integer 
modes in the NS sector, and half-integer modes in the R sector. The same is true, of 
course, for their right-moving counterparts. 

\subsubsection{Massless states in the NS/NS sector}
\label{NSNStwisted}
Since in the NS sector 
the fermionic coordinates along the directions 2 and 3 are periodic and possess 
zero-modes, the vacuum of the world-sheet theory of the fields $\Psi^{2}$ and 
$\Psi^3$ is degenerate and carries a representation of the 4$d$ Clifford algebra formed 
by their zero-modes. 
With respect to the SO(4) isometry 
of $\mathbb{C}_{(2)}\times \mathbb{C}_{(3)}$, these zero-modes build a 4$d$ Dirac 
spinor, which decomposes into its chiral and anti-chiral parts: $(\mathbf{2},\mathbf{1}) 
\oplus (\mathbf{1},\mathbf{2})$. Given our choice for the embedding of the $\mathbb{Z}_2$ 
action into SO(4), the anti-chiral part $(\mathbf{1},\mathbf{2})$ is not 
invariant under the orbifold and is projected out. Therefore, we just remain with the 
chiral spinor $(\mathbf{2},\mathbf{1})$, whose two components are labeled by an index 
$\alpha$. 
From the world-sheet point of view, this chiral spinor is created by a 4$d$ chiral spin 
field \cite{Friedan:1985ge,Kostelecky:1986xg}
\begin{equation}
S^\alpha(z)
\label{spinchiral}
\end{equation} 
which is a conformal field of weight 1/4.

Due to the twisted boundary conditions (\ref{twistcond}), the bosonic coordinates $Z^2$ 
and $Z^3$ along the orbifold do not possess zero-modes. The momentum can only be defined in 
the directions $Z^1$, $Z^4$ and $Z^5$ that have the standard behavior. 
We find it convenient to use a complex notation for the momentum analogous to the 
one in (\ref{coords}), and thus we define
\begin{equation}
\kappa_i=\frac{k_{2i-1} + \ii \,k_{2i}}{\sqrt{2}}\qquad\mbox{and}\qquad
\bar{\kappa}_{i}=\frac{k_{2i-1} - \ii \,k_{2i}}{\sqrt{2}}
\label{kappas}
\end{equation}
where $k$ is the momentum in real notation. Then, in the twisted sector, the usual 
plane-wave factor $:\!\rme^{\ii\,k\cdot X}\!:$ that appears in the vertex operators 
describing string exictations is written as follows
\begin{equation}
:\!\rme^{\ii\,\bar{\kappa}\cdot Z(z)+\ii\,\kappa\cdot \bar{Z}(z)}\!:
\label{planewave}
\end{equation}
where only $\kappa_1$, $\kappa_4$ and $\kappa_5$ (and their complex conjugates) 
are defined. The operator (\ref{planewave})
is a conformal field of weight $\kappa\cdot\bar{\kappa}=\frac{1}{2}k^2$.

Finally, to describe physical vertex operators in the standard $(-1)$-superghost picture 
of the NS sector, one introduces the vertex operator
\begin{equation}
:\!\rme^{-\phi(z)}\!:
\label{superghostNS}
\end{equation}
where $\phi(z)$ is the field appearing in the bosonization formulas of the superghost 
system \cite{Friedan:1985ge}. The operator (\ref{superghostNS}) is a conformal field of 
weight 1/2.

We have now all ingredients to construct a vertex operator that describes a 
physical left-moving excitation at the massless level in the NS twisted sector. 
This is obtained by taking the product of the twist field 
(\ref{Delta}), the spin field (\ref{spinchiral}), the plane-wave factor (\ref{planewave}) and 
the superghost term (\ref{superghostNS}). In this way we obtain\,%
\footnote{For simplicity, from now on in all vertex operators 
we will suppress the $:\,:$ notation, but the normal ordering will be always present.}
\begin{equation}
\mathcal{V}^\alpha(z)= \Delta(z)\,S^\alpha(z)\,\rme^{-\phi(z)}\,\rme^{\ii\,\bar{\kappa}
\cdot Z(z)+\ii\,\kappa\cdot \bar{Z}(z)}~,
\label{Val}
\end{equation}
which is a conformal field of weight 1 if  $\kappa\cdot\bar{\kappa}=\frac{1}{2}k^2=0$. 
In the following we will consider the closed strings as providing 
a constant background for the gauge theory, and thus in these vertex operators
we will set the momentum to zero. We also observe that the vertices (\ref{Val}) are 
preserved by the GSO projection of the NS sector. Indeed, the sum of the spinor weights 
minus the superghost charge is an even integer.

Exploiting the conformal properties of the various factors, it is easy to check that\,%
\footnote{Here and in the following, we understand the $\delta$-function enforcing 
momentum conservation.\label{footnotedelta}}
\begin{equation}
\big\langle \mathcal{V}^\alpha(z) \,\mathcal{V}^\beta(z')\big\rangle=
\frac{(\epsilon^{-1})^{\alpha\beta}}{(z-z')^2}\,,
\label{VVtilde}
\end{equation}
where 
\begin{equation}
\epsilon=\begin{pmatrix}
 0 & -1  \\
 +1 & 0 \\
\end{pmatrix}
\label{epsilon}
\end{equation}
is the chiral part of the charge conjugation matrix $\widehat{C}$ in four dimensions (see 
Appendix~\ref{twistedNSspinors} for details and our conventions).

The same construction goes through in the right-moving sector, where one finds the vertex 
operators
\begin{equation}
\widetilde{\mathcal{V}}^\alpha(\bar{z})
= \widetilde{\Delta}(\bar{z})\,\widetilde{S}^\alpha(\bar{z})\,
\rme^{-\widetilde{\phi}(\bar{z})}
\,\rme^{\ii\,\bar{\kappa}\cdot \widetilde{Z}(\bar{z})+\ii\,\kappa\cdot 
\widetilde{\bar{Z}}(\bar{z})}
\label{Var}
\end{equation}
which have the same form of the two-point function as in (\ref{VVtilde}) but with anti-holomorphic coordinate dependence.

Overall, the massless spectrum in the twisted NS/NS sector contains four states described 
by the vertices $\mathcal{V}^\alpha(z)\,\widetilde{\mathcal{V}}^\beta(\bar{z})$
in the $(-1,-1)$-superghost picture. 
The four independent components can be decomposed 
into a real scalar $b$ and a triplet $b_c$ (with $c=1,2,3$), transforming, respectively, in the
$(\mathbf{1},\mathbf{1})$ and $(\mathbf{3},\mathbf{1})$ representations of SO(4). They
correspond to the following vertex operators:
\begin{equation}
\begin{aligned}
b&~~\longleftrightarrow~~\cV_{b}(z,\bar{z}) = \ii\,\epsilon_{\alpha \beta}\,{\cV}^{\alpha}(z)
\,\widetilde{\cV}^{\beta}(\bar{z})~,\\
b_c&~~\longleftrightarrow~~
\cV_{b_c}(z,\bar{z})= (\epsilon\,\tau_c)_{\alpha \beta}\,{\cV}^{\alpha}(z)\,
\widetilde{\cV}^{\beta}(\bar{z})~,
\end{aligned}
\label{NSvertexops}
\end{equation}
where $\tau_c$ are the usual Pauli matrices.

\subsubsection{Massless states in the R/R sector}
\label{RRtwisted}
In the twisted R sector, the bosonic coordinates in the complex directions 2 and 3
have, of course, the same monodromy properties as in the NS sector, whereas the 
corresponding fermionic coordinates $\Psi^2$, $\Psi^3$ and their complex 
conjugates are anti-periodic.
This means that in those directions the world-sheet vacuum is non-degenerate.
On the contrary, the fermionic fields $\Psi^1$, $\Psi^4$ and $\Psi^5$ and their complex 
conjugates are periodic as usual in the R sector and possess zero-modes. Therefore the world-sheet vacuum in this twisted sector is degenerate and carries a 
representation of the 6$d$ Clifford algebra generated by the zero modes of the periodic 
fermions. These form a Dirac spinor of SO(6) which decomposes into a chiral part, 
transforming in the $\mathbf{4}$ of SO(6), plus an anti-chiral part transforming in the 
$\bar{\mathbf{4}}$.

From the world-sheet point of view, the chiral spinor is created by a 6$d$ chiral spin 
field \cite{Friedan:1985ge,Kostelecky:1986xg}
\begin{equation}
S^{A}(z)\,,
\label{SA}
\end{equation}
with $A$ taking four values. Likewise, the anti-chiral spinor corresponds to the
6$d$ anti-chiral spin field
\begin{equation}
S^{\dot{A}}(z) \,,
\label{SAdot}
\end{equation}
where also the dotted index $\dot{A}$ takes four values. Both $S^A$ and $S^{\dot{A}}$ 
are conformal fields of weight 3/8.

In the R sector there are two standard choices for the superghost picture:
the $(-\frac{1}{2})$-picture and the $(-\frac{3}{2})$-picture, created respectively by the 
operators
\begin{equation}
\rme^{-\frac{1}{2}\phi(z)}\qquad\mbox{and}\qquad
\rme^{-\frac{3}{2}\phi(z)}\,, 
\label{1232}
\end{equation}
which are both conformal fields of weight 3/8.

The GSO projection selects the combinations $S^A(z)\,\rme^{-\frac{1}{2}\phi(z)}$ and
$S^{\dot{A}}(z)\,\rme^{-\frac{3}{2}\phi(z)}$, for which the sum of the spinor weights 
minus the superghost charge is an even integer. Then, using these ingredients we can 
build the following vertex operators
\begin{subequations}
\begin{align}
\cV^{A}(z)&=\Delta(z)\,S^A(z)\,\rme^{-\frac{1}{2}\phi(z)}
\,\rme^{\ii\,\bar{\kappa}\cdot Z(z)+\ii\,
\kappa\cdot \bar{Z}(z)}\,~,
\label{VA12}\\
\cV^{\dot{A}}(z)&=\Delta(z)\,S^{\dot{A}}(z)\,\rme^{-\frac{3}{2}\phi(z)}
\,\rme^{\ii\,\bar{\kappa}\cdot Z(z)+\ii\,\kappa\cdot \bar{Z}(z)}~,
\label{VAdot32}
\end{align}
\label{Vs}%
\end{subequations}
which are conformal fields of dimension 1 
if $\kappa\cdot\bar{\kappa}=\frac{1}{2}k^2=0$.
{From} the conformal properties of the various components, it is easy to check that
\begin{equation}
\big\langle \mathcal{V}^A(z) \,\mathcal{V}^{\dot{B}}(z')\big\rangle=
\frac{(C^{-1})^{A\dot{B}}}{(z-z')^2} \,,
\label{VVtildeR}
\end{equation}
where $C$ is the charge conjugation matrix of  the spinor representations of SO(6) 
(see Appendix \ref{twistedRspinors}).

The same construction goes on in the right-moving sector, where one finds the vertex 
operators
\begin{subequations}
\begin{align}
\widetilde{\mathcal{V}}^A(\bar{z})
&= \widetilde{\Delta}(\bar{z})\,\widetilde{S}^A(\bar{z})\,\rme^{-\frac{1}{2}
\widetilde{\phi}(\bar{z})}
\,\rme^{\ii\,\bar{\kappa}\cdot \widetilde{Z}(\bar{z})+\ii\,\kappa\cdot 
\widetilde{\bar{Z}}(\bar{z})}~,
\label{VtildeA12}\\
\widetilde{\mathcal{V}}^{\dot{A}}(\bar{z})
&= \widetilde{\Delta}(\bar{z})\,\widetilde{S}^{\dot{A}}(\bar{z})\,
\rme^{-\frac{3}{2}\widetilde{\phi}(\bar{z})}\,
\rme^{\ii\,\bar{\kappa}\cdot \widetilde{Z}(\bar{z})+
\ii\,\kappa\cdot \widetilde{\bar{Z}}(\bar{z})}~,
\label{VtildeAdot32}
\end{align}
\label{Vtildes}%
\end{subequations}
which have the same two-point function as in (\ref{VVtildeR}).

Using the vertex operators (\ref{Vs}) and (\ref{Vtildes}) we can study the massless 
spectrum of the twisted R/R sector. In the asymmetric 
$(-\frac{1}{2},-\frac{3}{2})$-superghost picture the vertex operators 
$\mathcal{V}^A(z)\,\widetilde{\mathcal{V}}^{\dot{B}}(\bar{z})$ describe R/R 
potentials\,%
\footnote{As shown in \cite{Billo:1998vr} the full BRST invariant vertex operators 
describing the R/R potentials in the asymmetric supeghost picture are actually a sum of 
infinite terms with multiple insertions of superghost zero-modes. Here we only consider 
the first one of these terms, since all the others decouple from the physical amplitudes 
we will consider and thus can be neglected for our present purposes.} which have sixteen 
independent components. These can be decomposed into a scalar $c$ and 
a 2-index anti-symmetric tensor $c_{MN}$ of SO(6) that correspond
to the vertex operators
\begin{equation}
\begin{aligned}
c&~~\longleftrightarrow~~\cV_{c}(z,\bar{z}) 
= C_{A\dot{B}}\,{\cV}^{A}(z)\,\widetilde{\cV}^{\dot{B}}(\bar{z})~,\\
c_{MN}&~~\longleftrightarrow~~
\cV_{c_{MN}}(z,\bar{z})= (C\,\Gamma_{MN})_{A\dot{B}}\,{\cV}^{A}(z)\,
\widetilde{\cV}^{\dot{B}}(\bar{z})~,
\end{aligned}
\label{Rvertexops}
\end{equation}
where $\Gamma_{MN}=\frac{1}{2}[\Gamma_M,\Gamma_N]$, with $\Gamma_M$ being 
the Dirac matrices of SO(6) (see Appendix \ref{twistedRspinors}).

\section{Fractional D3-branes in the $\mathbb{Z}_2$ orbifold}		
\label{sec:fD3}

We engineer the 4$d$ gauge theory supporting the surface defect by means of fractional
D3-branes in the $\mathbb{Z}_2$ orbifold background (\ref{orbifold}). 
Differently from the case usually considered in the literature \cite{Douglas:1996sw}
in which the fractional D3-branes are entirely transverse to the orbifold, we take 
fractional D3-branes whose world-volume extends partially along the orbifold. In 
particular, using the notation introduced in the previous section, we consider D3-branes 
that extend along the complex directions 1 and 2, and are transverse to the complex 
directions 3, 4 and 5. Thus, the $\mathbb{Z}_2$ orbifold acts on one complex 
longitudinal and one complex transverse direction.
This fact has two important consequences: firstly, on the D3-brane world-volume, 
one finds the same content of massless fields as in 
$\mathcal{N}=4$ super Yang-Mills theory; 
secondly, since the orbifold acts only on 
one of the two complex directions of the world-volume, a 2$d$ surface
defect is naturally introduced in the gauge theory. 
Our goal is to show that this defect is precisely a GW monodromy defect.

To do so we first clarify the properties of the fractional D3-branes in the $\mathbb{Z}_2$ 
orbifold from the closed string point of view, using the boundary state formalism\,%
\footnote{For a review on the boundary state formalism, see for example 
\cite{DiVecchia:1999mal,DiVecchia:1999fje}.}, and then from the open string point of 
view by analyzing the  world-volume massless fields.

\subsection{Boundary states}
\label{bdrystate}

In a $\mathbb{Z}_2$ orbifold there are two types of fractional 
D-branes that correspond to the two irreducible representations of the orbifold group. We 
label these two types of D-branes by an index $I=0,1$. The D-branes with $I=0$ carry 
the trivial representation in which the $\mathbb{Z}_2$ element $g$ is represented 
by $+1$, while the D-branes with $I=1$ carry the other representation in which $g$ is 
represented by $-1$.
The two types of fractional branes therefore only differ by a sign in front of the twisted 
sectors. With this in mind, the fractional D3-branes can be represented in the boundary
state formalism in the following schematic way \cite{DiVecchia:1999mal,DiVecchia:1999fje,
Bertolini:2001gq}:
\begin{equation}
|\mathrm{D}3;I\rangle =\mathcal{N}\,|\mathrm{U}\rangle+
\mathcal{N}'\,|\mathrm{T};I\rangle
\quad\mbox{with}\quad |\mathrm{T};I\rangle=(-1)^{I}\,|\mathrm{T}\rangle~.
\label{fracD3s}
\end{equation}
Here $\mathcal{N}$ and $\mathcal{N}'$ are
dimensionful normalization factors related to the brane tension, and 
$|\mathrm{U}\rangle$ and $|\mathrm{T}\rangle$ are the untwisted and
twisted Ishibashi states that enforce the identification between the left and right moving 
modes in the untwisted and twisted sectors, respectively. For our purposes we do not
need to write the explicit expressions of these quantities which can be obtained by 
factorizing the 1-loop open-string partition function in the closed string channel, and 
thus we refer to the original literature and in particular to \cite{Bertolini:2001gq}, where 
also the case of D3-branes partially extending along the orbifold has been considered. 
However, for clarity, we recall the essential
information that will be needed in the following, namely that both $|\mathrm{U}\rangle$ 
and $|\mathrm{T}\rangle$ have a component in the NS/NS sector and a component 
in the R/R sector and that, after GSO projection, the twisted part of the boundary state is
\begin{equation}
|\mathrm{T}\rangle =|\mathrm{T}\rangle_{\mathrm{NS}}
+|\mathrm{T}\rangle_{\mathrm{R}}
\end{equation}
with
\begin{subequations}
\begin{align}
|\mathrm{T}\rangle_{\mathrm{NS}} &= (\widehat{C}{\gamma}_3
{\gamma}_4)_{\alpha\beta}
\,|\alpha\rangle\,|\widetilde{\beta}\rangle+ \cdots~,
\label{bdryTNS}\\[1mm]
|\mathrm{T}\rangle_{\mathrm{R}} &= ({C}\Gamma_1\Gamma_2)_{A\dot{B}}\,|A\rangle
\,|\widetilde{\dot{B}}\rangle+ \cdots~.
\label{bdryTR}
\end{align}
\label{bdryT}%
\end{subequations}
Here the kets represent the ground states created by acting on the untwisted vacuum
with the vertex operators (\ref{Val}), (\ref{Var}) of the NS/NS twisted sector and with the 
vertex operators (\ref{VA12}) and (\ref{VtildeAdot32}) of the R/R twisted sector, namely
\begin{subequations}
\begin{align}
|\alpha\rangle&=\lim_{z\to 0}\mathcal{V}^\alpha(z)\,|0\rangle~,\quad
|\widetilde{\beta}\rangle=\lim_{\bar{z}\to 0}\widetilde{\mathcal{V}}^\beta(\bar{z})\,
\widetilde{|0\rangle}~,
\label{bdryTNS1}\\[1mm]
|A\rangle&=\lim_{z\to 0}\mathcal{V}^A(z)\,|0\rangle~,\quad
|\widetilde{\dot{B}}\rangle=\lim_{\bar{z}\to 0}\widetilde{\mathcal{V}}^{\dot{B}}(\bar{z})
\,|\widetilde{0}\rangle~.
\label{bdryTR1}
\end{align}
\label{bdryT1}%
\end{subequations}
In (\ref{bdryT}) the ellipses stand for terms involving higher excited states which will not 
play any role in our analysis. We remark that the coefficient 
$(\widehat{C}\gamma_3\gamma_4)_{\alpha\beta}$ 
in the NS/NS component (\ref{bdryTNS}) is the appropriate one for our D3-branes since 
in the NS/NS twisted sector the ground states are spinors of the 4$d$ space spanned by 
the real coordinates $x_3$, $x_4$, $x_5$ and $x_6$, of which only the directions $x_3$ 
and $x_4$ are longitudinal to the D3-brane world-volume. 
Therefore the product of the SO(4) $\gamma$-matrices
${\gamma}_3{\gamma}_4$ must appear in the prefactor.
Notice that the GSO projection only selects the chiral block of the matrix 
$\widehat{C}{\gamma}_3{\gamma}_4$, as it is indicated by the undotted indices.
Likewise, in the R/R component (\ref{bdryTR}) the coefficient 
$({C}\Gamma_1\Gamma_2)_{A\dot{B}}$
is due to the fact that in R/R twisted sector the ground states are spinors in the 6$d$ 
space in which the real coordinates $x_1$ and $x_2$ belong to the D3-brane 
world-volume while the real coordinates $x_7$, $x_8$, $x_9$ and $x_{10}$ are 
transverse. This explains why the product of the SO(6) $\Gamma$-matrices
$\Gamma_1\Gamma_2$ appears in the prefactor. Again the GSO projection selects only the 
chiral/anti-chiral block of the matrix 
$C\Gamma_1\Gamma_2$, as indicated by the pair of undotted/dotted indices.

The boundary state $|\mathrm{D}3;I\rangle$ introduces a boundary on the closed string 
world-sheet along which the left and right moving modes are identified. As we explain in 
detail in Appendix~\ref{app:gluing}, from (\ref{bdryT}) one can derive that the
right moving parts of the twisted closed string vertex operators
are reflected on a boundary of type $I$ with the following rules
\begin{subequations}
\begin{align}
\widetilde{\mathcal{V}}^\alpha(\bar{z})&~\longrightarrow~~ (-1)^I
(\gamma_4\gamma_3)^\alpha_{~\beta}\,\mathcal{V}^\beta(\bar{z})~,\label{reflexNS}
\\[1mm]
\widetilde{\mathcal{V}}^{\dot{A}}(\bar{z})&~\longrightarrow~~ (-1)^I
(\Gamma_1\Gamma_2)^{\dot{A}}_{~\dot{B}}\,
\mathcal{V}^{\dot{B}}(\bar{z})~.\label{reflexR}
\end{align}
\label{reflex}%
\end{subequations}
These reflection rules will be important in computing closed string amplitudes involving 
twisted fields in the presence of the fractional D3-branes.

\subsection{The open string spectrum}
\label{openspectrum}
We now  analyze the spectrum of the massless excitations defined on the 
world-volume of the fractional D3-branes. For definiteness we take a fractional D3-brane of type 0, but of course completely 
similar considerations apply to a D3-brane of type 1. Since the world-volume extends in 
the first two complex directions
and the orbifold acts on the second one, it is convenient, as remarked at the start of Section \ref{subsec:md}, to distinguish the directions that are along and transverse to the orbifold. We will label the longitudinal 
variables (momentum, coordinates and so on) by a subscript 
$\parallel$, which involves the components along the first complex direction. We will similarly use the subscript $\perp$ to label the components along the second 
complex direction. 
The reason for these labels is that the first complex direction 
is longitudinal to the surface defect that the D3-branes realize, while the second 
direction is transverse to it. 
In particular, using the complex notation introduced
in Section~\ref{KTorbifold}, we define the combinations
\begin{equation}
\begin{aligned}
\kappa_\parallel \!\cdot\! Z_\parallel =\kappa_1\,\overbar{Z}^1+\overbar{\kappa}_1\,Z^1~,\\
\kappa_\perp \!\cdot\! Z_\perp =\kappa_2\,\overbar{Z}^2+\overbar{\kappa}_2\,Z^2~,
\end{aligned}
\end{equation}
and
\begin{equation}
\begin{aligned}
\kappa_\parallel\!\cdot\! \Psi_\parallel =\kappa_1\,\overbar{\Psi}^1+\overbar{\kappa}_1
\,\Psi^1~,\\
\kappa_\perp\!\cdot\!  \Psi_\perp =\kappa_2\,\overbar{\Psi}^2+\overbar{\kappa}_2\,\Psi^2~.
\end{aligned}
\end{equation}
Notice that under the $\mathbb{Z}_2$ orbifold parity, $\kappa_\parallel \!\cdot\!  
Z_\parallel $ and $\kappa_\parallel\!\cdot\!  \Psi_\parallel$ are even, while 
$\kappa_\perp \!\cdot\!  Z_\perp$ and $\kappa_\perp\!\cdot\!  \Psi_\perp$
are odd.

Let us consider the bosonic NS sector. 
In the familiar case when the D3-branes are completely transverse to the orbifold, the 
gauge vector field $A_\mu$ is typically represented in the $(0)$-superghost picture by 
the standard vertex operator
\begin{equation}
\big(\ii\,\partial X^\mu+k\cdot\psi\,\psi^\mu\big)\,\rme^{\ii\,k\cdot X}~.
\label{VAmu}
\end{equation}
In our case things are different. First of all, in the plane wave factor 
$\rme^{\ii\,k\cdot X}$ we have to distinguish the parallel and perpendicular parts which 
behave differently under the orbifold, and thus we are naturally led to consider the 
following structures
\begin{equation}
\begin{aligned}
\cos(\kappa_\perp \!\cdot\! Z_\perp)\,\rme^{\ii\,\kappa_\parallel\cdot Z_\parallel}~,\\
\ii\,\sin(\kappa_\perp \!\cdot\!  Z_\perp)\,\rme^{\ii\,\kappa_\parallel\cdot Z_\parallel}~,
\end{aligned}
\label{sincos}
\end{equation}
which are respectively even and odd under $\mathbb{Z}_2$. Therefore, they can be 
combined with other even and odd structures to make invariant vertex operators selected
by the orbifold projection. Similarly, also the $k\cdot\psi$ combination appearing in 
(\ref{VAmu}) has to be split into a parallel and a perpendicular component.

Applying these considerations, it is not difficult to realize that the gauge field $A_1$ 
along  the parallel directions is described by the following vertex operator in the 
$(0)$-superghost picture\,%
\footnote{We consider the $(0)$-superghost picture since it is the relevant one for the 
applications discussed in Section~\ref{correlators}, but of course our analysis can be 
done also in any other superghost picture of the NS sector.}
\begin{equation}
A_1~\longrightarrow~\mathcal{V}_{A_1}=\Big[
\big(\ii\,\partial{Z}^1+\kappa_\parallel\!\cdot\! \Psi_\parallel\,{\Psi}^1\big)
\cos(\kappa_\perp \!\cdot\!  Z_\perp)+\ii\,
\kappa_\perp\!\cdot\! \Psi_\perp\,{\Psi}^1\,\sin(\kappa_\perp \!\cdot\!  Z_\perp)\Big]
\rme^{\ii\,\kappa_\parallel\cdot Z_\parallel}~.
\label{VA1}
\end{equation}
Each term in this expression is invariant under $\mathbb{Z}_2$. 
For instance, the terms $\ii\,\partial {Z}^{1}$ or 
$\kappa_\parallel\!\cdot\! \Psi_\parallel\,{\Psi}_1$, which are 
$\mathbb{Z}_2$-even, are multiplied with the cosine combination 
$\cos(\kappa_\perp Z_\perp)$ which is also even, so that the product is invariant under 
the orbifold action. Similarly, the odd term 
$\kappa_\perp\!\cdot\! \Psi_\perp\,{\Psi}^1$ is multiplied by the sine combination 
$\sin(\kappa_\perp \!\cdot\! Z_\perp)$, which is also odd, to make an even expression 
under $\mathbb{Z}_2$.
Furthermore, it is easy to check that $\mathcal{V}_{A_1}$ is a conformal field of weight 1 
if $\kappa\cdot\bar{\kappa}=\frac{1}{2}k^2=0$. The vertex operator for the complex conjugate component $\overbar{A}_1$ of the gauge field is simply obtained by replacing $\partial Z^1$ with $\partial \overbar{Z}^1$ and 
$\Psi^1$ with $\overbar{\Psi}^1$ in the above expression.

The gauge field $A_2$ along the second complex direction of the D3-brane 
world-volume is instead described by the following vertex operator
\begin{equation}
A_2~\longrightarrow~\mathcal{V}_{A_2}=\Big[
\big(\ii\,\partial{Z}^2+\kappa_\parallel\!\cdot\! \Psi_\parallel\,{\Psi}^2\big)\,
\ii\,\sin(\kappa_\perp \!\cdot\!  Z_\perp)+
\kappa_\perp\!\cdot\!  \Psi_\perp\,{\Psi}^2\,\cos(\kappa_\perp \!\cdot\!  Z_\perp)\Big]
\rme^{\ii\,\kappa_\parallel\cdot Z_\parallel}~.
\label{VA2}
\end{equation}
Notice that the position of the cosine and sine combinations is different with respect 
to (\ref{VA1}), but this is precisely what is needed to obtain an invariant vertex in this 
case. Again this vertex is a conformal field of weight 1 if the field is massless. 
The operator describing the complex conjugate component $\bar{A}_2$ is obtained by 
replacing  $\partial Z^2$ with $\partial \overbar{Z}^2$ and $\Psi^2$ with $\overbar{\Psi}^2$ 
in (\ref{VA2}).

Let us now consider the massless scalar fields. Without the orbifold, on the D3-brane 
world-volume there are three complex scalars that together with the gauge vector 
provide the bosonic content of the $\mathcal{N}=4$ vector multiplet. When the orbifold 
acts entirely in the transverse directions, only one of these scalars remains in the 
invariant spectrum, thus reducing the supersymmetry 
from $\mathcal{N}=4$ to $\mathcal{N}=2$. In our case, instead, when the orbifold 
acts partially along the world-volume, all three complex scalars remain. Denoting them
by $\Phi$ and $\Phi_r$ with $r=4,5$, they are described by the following three 
$\mathbb{Z}_2$-invariant vertices
\begin{equation}
\Phi~\longrightarrow~\mathcal{V}_{\Phi}=\Big[
\big(\ii\,\partial{Z}^3+\kappa_\parallel\!\cdot\!  \Psi_\parallel\,{\Psi}^3\big)\,
\ii\,\sin(\kappa_\perp\!\cdot\!  Z_\perp)+
\kappa_\perp\!\cdot\! \Psi_\perp\,{\Psi}^3\,\cos(\kappa_\perp \!\cdot\!  Z_\perp)\Big]
\rme^{\ii\,\kappa_\parallel\cdot Z_\parallel}~.
\label{VPhi}
\end{equation}
and 
\begin{equation}
\Phi_r~\longrightarrow~\mathcal{V}_{\Phi_r}=\Big[
\big(\ii\,\partial{Z}^r+\kappa_\parallel\!\cdot\! \Psi_\parallel\,{\Psi}^r\big)
\cos(\kappa_\perp\!\cdot\!  Z_\perp)+\ii\,
\kappa_\perp\!\cdot\!  \Psi_\perp\,{\Psi}^r\,\sin(\kappa_\perp \!\cdot\!  Z_\perp)\Big]
\rme^{\ii\,\kappa_\parallel\cdot Z_\parallel}~.
\label{VPhii}
\end{equation}
Since the scalars are massless, these vertices are conformal operators of weight 1.

A similar analysis can be repeated also for the fermionic R sector, where one can find 
sixteen massless fermions that are the supersymmetric partners of the bosonic fields 
listed above. 

In conclusion we see that when the fractional D3-branes extend partially along the 
orbifold, the latter does not project the open string spectrum by removing some 
excitations, as it does when the fractional D3-branes are totally transverse, but instead it 
reorganizes the fields in such a way that they behave differently along the $\parallel$ 
and $\perp$ subspaces into which the 4$d$ world-volume of the D3-branes is divided. Another piece of evidence for the defect interpretation is the 1-loop open string partition function \cite{Bertolini:2001gq}, which receives contributions both from modes that propagate in all four dimensions of the world-volume and also from modes that propagate only in the $\parallel$ subspace. This is precisely what one expects for a surface defect in the 4$d$ gauge theory, extended along the $\parallel$ subspace. Moreover, if we consider a system 
made of $n_0$ fractional D3-branes of type 0 and $n_1$ fractional D3-branes of type 1, 
we engineer a 4$d$ theory with gauge group U($n_0+n_1$) broken to the Levi group 
U($n_0$)$\times$U($n_1)$ at the orbifold fixed plane. In the case of special unitary groups, the overall U(1) factor has to be removed.

\section{Open/closed correlators}
\label{correlators}
Our next step is to show that there are non-vanishing interactions between the twisted 
closed string sectors discussed in Section~\ref{KTorbifold} and the open string fields 
 introduced in the previous section. In particular, 
we will show that there are non-vanishing amplitudes corresponding to the  
diagram represented in (\ref{tadpole}). The reason why such open/closed amplitudes exist 
is that a D-brane inserts a boundary in the closed string world-sheet along which
the left- and right-moving modes are identified. Thus, the two components of the 
closed string vertex operators effectively behave as two open string vertices which 
can have a non-vanishing interaction with a third open string vertex operator describing 
an excitation of the gauge theory on the brane word-volume. 
In the following we are going to systematically compute these 
open/closed string amplitudes, starting from the twisted NS/NS sector.

\subsection{Correlators with NS/NS twisted fields}
As we discussed in the Section~\ref{NSNStwisted}, in the twisted NS/NS sector the 
fermionic fields in the 4$d$ space where the $\mathbb{Z}_2$
orbifold acts have zero modes that build a spinor representation of SO(4). A fractional 
D3-brane that partially extends along the orbifold breaks this SO(4) into SO(2)$\times$SO(2).
In this breaking, the singlet $b$ remains, while the triplet $b_c 
\in (\mathbf{3},\mathbf{1})$ decomposes 
into a scalar $b'$ and a doublet $b_\pm$ of complex conjugate fields. 
The vertex operators corresponding to these four fields 
can be read from (\ref{NSvertexops}), which we rewrite here for convenience
\begin{subequations}
\begin{align}
b &~~\longleftrightarrow~~\mathcal{V}_b(z,\bar{z})=
\ii\,\epsilon_{\alpha\beta}\,\mathcal{V}^\alpha(z)\,\widetilde{V}^\beta(\bar{z})~,
\label{Vxi}\\
b' &~~\longleftrightarrow~~\mathcal{V}_{b'}(z,\bar{z})=
(\epsilon\tau_3)_{\alpha\beta}\,\mathcal{V}^\alpha(z)\,
\widetilde{V}^\beta(\bar{z})~,
\label{Vxiprime}\\
b_\pm&~~\longleftrightarrow~~\mathcal{V}_{b_\pm}(z,\bar{z})=
(\epsilon\tau_\pm)_{\alpha\beta}
\,\mathcal{V}^\alpha(z)\,\widetilde{V}^\beta(\bar{z})
\label{Vxipm}
\end{align}
\label{NSvertexops1}%
\end{subequations}
where $\tau_\pm=(\tau_1\pm\ii\,\tau_2)/2$.
Since we are going to regard the closed string fields as a background for the open string 
excitations, in all vertices (\ref{NSvertexops1}) we set the momentum to zero.

\subsubsection*{Correlators with $b$}
We begin by evaluating the couplings of the massless open string fields of a 
fractional D3-brane of type $I$ with the scalar $b$. These are given by 
\begin{equation}
\big\langle \mathcal{V}_{\text{open}} \big\rangle_{b;I}
=b\int\frac{dz\,d\bar{z}\,dx}{dV_{\text{proj}}}~\big\langle
\mathcal{V}_b(z,\bar{z})\,\mathcal{V}_{\text{open}}(x)\big\rangle_I
\label{VopenI}
\end{equation}
where $\mathcal{V}_{\text{open}}$ stands for any of the vertex operators described in
Section~\ref{openspectrum} and
\begin{equation}
\label{proj_vol}
dV_{\text{proj}} = \frac{dz\,d\bar{z}\,dx}{(z-\bar{z})(\bar{z}-x)(x-z)}
\end{equation}
is the projective invariant volume element. In (\ref{VopenI}) the integrals are performed
on the string word-sheet. In particular $z$ and $\bar{z}$, where the close string 
vertex operator is inserted, are points in the upper and lower half complex plane, 
respectively, while $x$ is a point on the real axis from which the open string is emitted. 
The integrand of (\ref{VopenI}) is
\begin{equation}
\begin{aligned}
\big\langle
\mathcal{V}_b(z,\bar{z})\,\mathcal{V}_{\text{open}}(x)\big\rangle_I&=\ii\,
\epsilon_{\alpha\beta}\,\big\langle\mathcal{V}^\alpha(z)\,
\widetilde{\mathcal{V}}^\beta(\bar{z})\,\mathcal{V}_{\text{open}}(x)\big\rangle_I 
\\[1mm]
&=
(-1)^I\,\ii\,\epsilon_{\alpha\beta}(\gamma_4\gamma_3)^\beta_{~\gamma}
\,\big\langle\mathcal{V}^\alpha(z)\,
\mathcal{V}^\gamma(\bar{z})\,\mathcal{V}_{\text{open}}(x)\big\rangle
\end{aligned}
\label{mixed}
\end{equation}
where the second line follows from the reflection rules (\ref{reflexNS}). Our task is 
therefore to compute the three-point functions $\big\langle\mathcal{V}^\alpha(z)\,
\mathcal{V}^\gamma(\bar{z})\,\mathcal{V}_{\text{open}}(x)\big\rangle$ for the various 
open string fields. 

Let us start with the components of the gauge field that are longitudinal to the defect. 
These are described by the vertex operator (\ref{VA1}). Factorizing 
the resulting amplitude in a product of correlation functions for the independent 
conformal fields, we find
\begin{align}
\big\langle\mathcal{V}^\alpha(z)\,
\mathcal{V}^\gamma(\bar{z})&\,\mathcal{V}_{A_1}(x)\big\rangle=
\big\langle\rme^{-\phi(z)}\,\rme^{-\phi(\bar{z})}\big\rangle\\[1mm]
&\times\Big[\ii\,\big\langle\partial{Z}_1(x)\rme^{\ii\,\kappa_\parallel\cdot
Z_\parallel(x)}\big\rangle \big\langle \Delta(z)
\Delta(\bar{z})\cos(k_\perp\!\cdot\! Z_\perp)(x)\big\rangle
\big\langle S^\alpha(z) S^\gamma(\bar{z})\big\rangle\notag\\[1mm]
&+\big\langle \rme^{\ii\,\kappa_\parallel\cdot Z_\parallel(x)}\big\rangle
\big\langle \Delta(z)\Delta(\bar{z})\cos(k_\perp\!\cdot\! Z_\perp)(x)\big\rangle
\big\langle S^\alpha(z) S^\gamma(\bar{z})\big\rangle
\big\langle\kappa_\parallel\!\cdot\!\Psi_\parallel(x){\Psi}_1(x)\big\rangle
\notag\\[1mm]
&+\ii\,\big\langle \rme^{\ii\,\kappa_\parallel\cdot Z_\parallel(x)}\big\rangle
\big\langle \Delta(z)\Delta(\bar{z})\,\sin(k_\perp\!\cdot\! Z_\perp)(x)\big\rangle
\big\langle S^\alpha(z) S^\gamma(\bar{z})\,\kappa_\perp\!\cdot\!\Psi_\perp(x) 
\big\rangle \big\langle{\Psi}_1(x)\big\rangle\Big]~.\notag
\end{align}
It is not difficult to realize that in each of the three lines in square brackets, there is always 
one factor that vanishes due to normal ordering. 
For example, in the first line it is the term containing 
$\ii\,\partial{Z}_1$ that vanishes, while in the second and third line it is the last factor 
involving the fermionic field $\Psi_1$ that gives zero. Therefore, 
\begin{equation}
\big\langle\mathcal{V}^\alpha(z)\,
\mathcal{V}^\gamma(\bar{z})\,\mathcal{V}_{A_1}(x)\big\rangle=0~,
\label{vvA1}
\end{equation}
 so that
\begin{equation}
\big\langle \mathcal{V}_{A_1} \big\rangle_{b;I}=0~.
\end{equation}

Let us now consider the components of the gauge field that are transverse to the defect. 
Using the corresponding vertex operator (\ref{VA2}), we obtain
\begin{align}
\big\langle\mathcal{V}^\alpha(z)\,
\mathcal{V}^\gamma(\bar{z})&\,\mathcal{V}_{A_2}(x)\big\rangle=
\big\langle\rme^{-\phi(z)}\,\rme^{-\phi(\bar{z})}\big\rangle\big\langle \rme^{\ii\,
\kappa_\parallel\cdot Z_\parallel(x)}\big\rangle\\[1mm]
&\times\Big[-\big\langle \Delta(z)
\Delta(\bar{z})\,\partial{Z}_2(x)\,\sin(k_\perp\!\cdot\! Z_\perp)(x)\big\rangle
\big\langle S^\alpha(z) S^\gamma(\bar{z})\big\rangle\notag\\[1mm]
&+\ii\,
\big\langle \Delta(z)\Delta(\bar{z})\,\sin(k_\perp\!\cdot\! Z_\perp)(x)\big\rangle
\big\langle\kappa_\parallel\!\cdot\!\Psi_\parallel(x)\big\rangle
\big\langle S^\alpha(z) S^\gamma(\bar{z}){\Psi}_2(x)\big\rangle\notag\\[1mm]
&+
\big\langle \Delta(z)\Delta(\bar{z})\cos(k_\perp\!\cdot\! Z_\perp)(x)\big\rangle
\big\langle S^\alpha(z) S^\gamma(\bar{z})\,\kappa_\perp\!\cdot\!\Psi_\perp(x) 
{\Psi}_2(x)\big\rangle\Big]~.\notag
\end{align}
As before, in the first and second lines inside the square brackets there are vanishing 
factors; instead, the third line is not zero and we remain with
\begin{equation}
\begin{aligned}
\big\langle\mathcal{V}^\alpha(z)\,
\mathcal{V}^\gamma(\bar{z})\,\mathcal{V}_{A_2}(x)\big\rangle&=
\big\langle\rme^{-\phi(z)}\,\rme^{-\phi(\bar{z})}\big\rangle\big\langle \rme^{\ii\,
\kappa_\parallel\cdot Z_\parallel(x)}\big\rangle
\big\langle \Delta(z)\Delta(\bar{z})\cos(k_\perp\!\cdot\! Z_\perp)(x)\big\rangle
\\[1mm]
&\qquad
\times\big\langle S^\alpha(z) S^\gamma(\bar{z})\,\kappa_\perp\!\cdot\!\Psi_\perp(x) 
{\Psi}_2(x)\big\rangle~.
\end{aligned}
\label{xiA2}
\end{equation}
Each correlator in this expression can be easily evaluated using standard conformal field 
theory methods; in particular we have
\begin{subequations}
\begin{align}
\big\langle\rme^{-\phi(z)}\,\rme^{-\phi(\bar{z})}\big\rangle &=\frac{1}{z-\bar{z}}~,
\label{superghost}\\[1mm]
\big\langle \rme^{\ii\,\kappa_\parallel\cdot Z_\parallel(x)}\big\rangle &=
\delta^{(2)}(\kappa_\parallel)~,\label{deltakappa}\\[1mm]
\big\langle \Delta(z)\Delta(\bar{z})\cos(k_\perp\!\cdot\! Z_\perp)(x)\big\rangle &=
\frac{1}{(z-\bar{z})^{\frac{1}{2}}}~,\label{deltadelta}\\[1mm]
\big\langle S^\alpha(z) S^
\gamma(\bar{z})\,\psi_m(x)\psi_n(x)\big\rangle
&=\frac{1}{2}\,\frac{(\gamma_n\gamma_m\widehat{C}^{-1})^{\alpha
\gamma}}{(z-\bar{z})^{-\frac{1}{2}}(z-x)(\bar{z}-x)}~.
\label{sspsipsi}
\end{align}
\label{CFT}%
\end{subequations}
The last correlator implies that
\begin{equation}
\begin{aligned}
\big\langle S^\alpha(z) S^\gamma(\bar{z})\,\kappa_\perp\!\cdot\!\Psi_\perp(x) 
{\Psi}_2(x)\big\rangle &= \ii\,\kappa_2\,\big\langle S^\alpha(z) S^
\gamma(\bar{z})\,\psi_3(x)\psi_4(x)\big\rangle\\
&=\ii\,\frac{\kappa_2}{2}\,\frac{(\gamma_4\gamma_3\widehat{C}^{-1})^{\alpha
\gamma}}{(z-\bar{z})^{-\frac{1}{2}}(z-x)(\bar{z}-x)}~.
\end{aligned}
\end{equation}
Putting everything together, we obtain
\begin{equation}
\big\langle\mathcal{V}^\alpha(z)\,
\mathcal{V}^\gamma(\bar{z})\,\mathcal{V}_{A_2}(x)\big\rangle
=\ii\,\frac{{\kappa}_2}{2}\,\frac{(\gamma_4\gamma_3\widehat{C}^{-1})^{\alpha
\gamma}}{(z-\bar{z})(z-x)(\bar{z}-x)}\,\delta^{(2)}(\kappa_\parallel)~.
\label{vvA2}
\end{equation}
Inserting this result in (\ref{mixed}) and performing the corresponding 
$\gamma$-matrix algebra, in the end we find
\begin{equation}
\big\langle \mathcal{V}_{A_2} \big\rangle_{b;I}
=(-1)^{I+1}\,b\,{\kappa}_2\,\delta^{(2)}(\kappa_\parallel)~.
\label{VA2fin}
\end{equation}
As is clear from this expression, the momentum conservation occurs only in 
the longitudinal directions, whereas the transverse momenta 
$\kappa_2$ and $\bar{\kappa}_2$ can be arbitrary. 
This fact implies that (\ref{VA2fin}) can be interpreted as a tadpole-like source for the 
gauge field $A_2$ which acquires a non-trivial profile in the transverse space.
We will explicitly compute this profile in the following section.

The calculation of the couplings of $b$ with the complex scalar $\Phi$ gauge theory 
proceeds along the same lines. One finds that the only non-vanishing contribution to the 
correlation function is given by
\begin{equation}
\begin{aligned}
\big\langle\mathcal{V}^\alpha(z)\,
\mathcal{V}^\gamma(\bar{z})\,\mathcal{V}_{\Phi}(x)\big\rangle&=
\big\langle\rme^{-\phi(z)}\,\rme^{-\phi(\bar{z})}\big\rangle\big\langle \rme^{\ii\,
\kappa_\parallel\cdot Z_\parallel(x)}\big\rangle
\big\langle \Delta(z)\Delta(\bar{z})\cos(k_\perp\!\cdot\! Z_\perp)(x)\big\rangle
\\[1mm]
&\qquad
\times\big\langle S^\alpha(z) S^\gamma(\bar{z})\,\kappa_\perp\!\cdot\!\Psi_\perp(x) 
{\Psi}_3(x)\big\rangle~.
\end{aligned}
\label{xiphi}
\end{equation}
The last factor is easily computed using (\ref{sspsipsi}) with the result
\begin{equation}
\begin{aligned}
\big\langle S^\alpha(z) S^\gamma(\bar{z})\,\kappa_\perp\!\cdot\!\Psi_\perp(x) 
{\Psi}_3(x)\big\rangle&=\frac{\kappa_2}{4}\,
\frac{\big((\gamma_5+\ii\,\gamma_6)(\gamma_3-\ii\,\gamma_4)
\widehat{C}^{-1}\big)^{\alpha
\gamma}}{(z-\bar{z})^{-\frac{1}{2}}(z-x)(\bar{z}-x)}
\\
&~+\frac{\bar{\kappa}_2}{4}\,
\frac{\big((\gamma_5+\ii\,\gamma_6)(\gamma_3+\ii\,\gamma_4)
\widehat{C}^{-1}\big)^{\alpha
\gamma}}{(z-\bar{z})^{-\frac{1}{2}}(z-x)(\bar{z}-x)}~.
\end{aligned}
\end{equation}
This implies that
\begin{equation}
\begin{aligned}
\big\langle\mathcal{V}^\alpha(z)\,
\mathcal{V}^\gamma(\bar{z})\,\mathcal{V}_{\Phi}(x)\big\rangle=
\frac{\Big[\big(
\frac{\kappa_2}{4}\,(\gamma_5+\ii\,\gamma_6)(\gamma_3-\ii\,\gamma_4)+
\frac{\bar{\kappa}_2}{4}\,(\gamma_5+\ii\,\gamma_6)(\gamma_3+\ii\,\gamma_4)\big)
\,\widehat{C}^{-1}\Big]^{\alpha\gamma}}{(z-\bar{z})(z-x)(\bar{z}-x)}
\,\delta^{(2)}(\kappa_\parallel)~.
\end{aligned}
\label{vvphi}
\end{equation}
When we plug this expression into (\ref{mixed}) and perform the resulting 
$\gamma$-matrix algebra we get zero, so that
\begin{equation}
\big\langle \mathcal{V}_{\Phi} \big\rangle_{b;I}=0
\label{xiPhifin}
\end{equation}
Finally, considering the scalars $\Phi_r$, we find that
\begin{equation}
\big\langle\mathcal{V}^\alpha(z)\,
\mathcal{V}^\gamma(\bar{z})\,\mathcal{V}_{\Phi_r}(x)\big\rangle=0
\label{vvPhir}
\end{equation}
since, like for $A_1$, the resulting correlator always contains a vanishing factor. Therefore,
\begin{equation}
\big\langle \mathcal{V}_{\Phi_r} \big\rangle_{b;I}=0~.
\end{equation}

\subsubsection*{Correlators with $b'$}
Let us now consider the couplings with the twisted scalar $b'$ whose vertex operator 
(\ref{Vxiprime}) has the polarization $\epsilon\tau_3$. The vanishing of
the correlators (\ref{vvA1}) and (\ref{vvPhir}) shows that $\mathbf{A}_1$
and $\mathbf{\Phi}_r$ do not couple to any NS/NS twisted field, including $b'$.
Also the non-vanishing correlators (\ref{vvA2}) and (\ref{vvphi}) give a zero
result for $b'$ due to the $\gamma$-matrix algebra. 
Therefore the field $b'$ does not couple to any of the massless open string fields of the 
gauge theory:  
\begin{equation}
\big\langle \mathcal{V}_{A_1} \big\rangle_{b';I}
=\big\langle \mathcal{V}_{A_2} \big\rangle_{b';I}
=\big\langle \mathcal{V}_{\Phi} \big\rangle_{b';I}
=\big\langle \mathcal{V}_{\Phi_r} \big\rangle_{b';I}=0~.
\end{equation}

\subsubsection*{Correlators with $b_\pm$}
The couplings of the doublet $b_\pm$ with the open string fields can be computed
along the same lines. We simply have to use the correlators (\ref{vvA1}), (\ref{vvA2}), 
(\ref{vvphi}) and (\ref{vvPhir}) and the polarizations $(\epsilon\tau_\pm)$ 
corresponding to $b_\pm$. Proceeding in this way we find
\begin{equation}
\big\langle \mathcal{V}_{A_1} \big\rangle_{b_\pm;I}
=\big\langle \mathcal{V}_{A_2} \big\rangle_{b_\pm;I}=
\big\langle \mathcal{V}_{\Phi} \big\rangle_{b_-;I}
=\big\langle \mathcal{V}_{\Phi_r} \big\rangle_{b_\pm;I}=0~.
\end{equation}
The vanishing of the coupling of $A_2$ with $b_\pm$ and of the coupling of
$\Phi$ with $b_-$ is again due to the structure of the resulting combinations of 
$\gamma$-matrices which have a vanishing trace. On the other hand, the terms
proportional to $\bar{\kappa}_2$ in (\ref{vvphi}) yield a non-zero result when
contracted with the polarization of $b_+$, leading to
\begin{equation}
\begin{aligned}
\big\langle \mathcal{V}_{\Phi} \big\rangle_{b_+;I}&=(-1)^{I+1}\,\ii\,b_+\,\bar{\kappa}_2\,
\delta^{(2)}(\kappa_\parallel)~.
\label{VPhiplus}
\end{aligned}
\end{equation}

\subsection{Correlators with R/R twisted fields}

Let us now consider the twisted fields of the R/R sector that we discussed in 
Section~\ref{RRtwisted}. In the twisted R/R sector the fermionic fields possess zero 
modes in the six dimensions that are orthogonal to the $\mathbb{Z}_2$ orbifold. They
realize spinor representations of SO(6), but when a fractional D3-brane is inserted, 
this group is broken to SO(2)$\times$SO(4). We are interested in giving a constant background value to some scalars that remain after this breaking. The scalar $c$ obviously remains, while 
the anti-symmetric tensor $c_{MN}\in \mathbf{15}$ decomposes
in various representations of the unbroken subgroup. In particular, we will consider only 
the component $c_{12}$ which is a scalar of SO(2)$\times$SO(4) that we denote $c'$. 
The vertex operators corresponding to $c$ and $c'$ are given in (\ref{Rvertexops}) which we rewrite here for convenience:
\begin{subequations}
\begin{align}
c&~~\longleftrightarrow~~\cV_c(z,\bar{z}) 
= C_{A\dot{B}}\,{\cV}^{A}(z)\,\widetilde{\cV}^{\dot{B}}(\bar{z})~, \label{Vc}
\\
c'&~~\longleftrightarrow~~
\cV_{c'}(z,\bar{z})= (C\,\Gamma_{12})_{A\dot{B}}\,{\cV}^{A}(z)\,
\widetilde{\cV}^{\dot{B}}(\bar{z})~.
\label{Vcprime}
\end{align}
\label{Rvertexops1}%
\end{subequations}
Again we take these vertices at zero momentum since we want to regard the closed 
string fields as a constant background.

\subsubsection*{Correlators with $c$}
The mixed correlators between the R/R twisted scalar $c$ and the open string massless
fields of a D3-brane of type $I$ are given by
\begin{equation}
\big\langle \mathcal{V}_{\text{open}} \big\rangle_{c;I}
=c\int\frac{dz\,d\bar{z}\,dx}{dV_{\text{proj}}}~\big\langle
\mathcal{V}_{c}(z,\bar{z})\,\mathcal{V}_{\text{open}}(x)\big\rangle_I
\label{VopenIC}
\end{equation}
with
\begin{equation}
\begin{aligned}
\big\langle
\mathcal{V}_{c}(z,\bar{z})\,\mathcal{V}_{\text{open}}(x)\big\rangle_I&=
C_{A\dot{B}}\,\big\langle\mathcal{V}^A(z)\,
\widetilde{\mathcal{V}}^{\dot{B}}(\bar{z})\,\mathcal{V}_{\text{open}}(x)\big\rangle_I 
\\[1mm]
&=
(-1)^I\,C_{A\dot{B}}(\Gamma_1\Gamma_2)^{\dot{B}}_{~\dot{C}}
\,\big\langle\mathcal{V}^A(z)\,
\mathcal{V}^{\dot{C}}(\bar{z})\,\mathcal{V}_{\text{open}}(x)\big\rangle~.
\end{aligned}
\label{mixedR}
\end{equation}
where the last step follows from the reflection rules (\ref{reflexR}).

The first coupling we consider is the one with the gauge field $A_1$. 
Using the vertex operator (\ref{VA1}) we find that there is only a single structure contributing
to the amplitude, namely 
\begin{equation}
\begin{aligned}
\big\langle\mathcal{V}^A(z)\,\mathcal{V}^{\dot{C}}(\bar{z})\,\mathcal{V}_{A_1}(x)
\big\rangle &=
\big\langle\rme^{-\frac{1}{2}\phi(z)}\,\rme^{-\frac{3}{2}\phi(\bar{z})}\big\rangle
\big\langle \rme^{\ii\,\kappa_\parallel\cdot Z_\parallel(x)}\big\rangle
\big\langle \Delta(z)\Delta(\bar{z})\cos(k_\perp\!\cdot\! Z_\perp)(x)\big\rangle
\\[1mm]
&\qquad
\times\big\langle S^A(z) S^{\dot{C}}(\bar{z})\,\kappa_\parallel\!\cdot\!\Psi_\parallel(x) 
{\Psi}_1(x)\big\rangle~.
\end{aligned}
\label{Cphi}
\end{equation}
The second and third factors are given in (\ref{deltakappa}) and (\ref{deltadelta}), while 
the other factors are obtained from the standard conformal field theory results,  namely
\begin{subequations}
\begin{align}
\big\langle\rme^{-\frac{1}{2}\phi(z)}\,\rme^{-\frac{3}{2}\phi(\bar{z})}\big\rangle
&=\frac{1}{(z-\bar{z})^{\frac{3}{4}}}~,\\[1mm]
\big\langle S^A(z) S^{\dot{C}}(\bar{z})\,\psi_M(x)\psi_N(x)\big\rangle
&=\frac{1}{2}\,\frac{(\Gamma_M\Gamma_N C^{-1})^{A\dot{C}}}
{(z-\bar{z})^{-\frac{1}{4}}(z-x)(\bar{z}-x)}~.
\end{align}
\end{subequations} 
The last correlator implies that
\begin{equation}
\begin{aligned}
\big\langle S^A(z) S^{\dot{C}}(\bar{z})\,\kappa_\parallel\!\cdot\!\Psi_\parallel(x) 
{\Psi}_1(x)\big\rangle &= \ii\,\kappa_1\,\big\langle S^A(z) 
S^{\dot{C}}(\bar{z})\,\psi_1(x)\psi_2(x)\big\rangle\\
&=\ii\,\frac{\kappa_1}{2}\,\frac{(\Gamma_1\Gamma_2C^{-1})^{A\dot{C}}}
{(z-\bar{z})^{-\frac{1}{4}}(z-x)(\bar{z}-x)}
\end{aligned}
\end{equation}
so that from (\ref{Cphi}) we get
\begin{equation}
\big\langle\mathcal{V}^A(z)\,\mathcal{V}^{\dot{C}}(\bar{z})\,\mathcal{V}_{A_1}(x)
\big\rangle=
\ii\,\frac{\kappa_1}{2}\,\frac{(\Gamma_1\Gamma_2C^{-1})^{A\dot{C}}}
{(z-\bar{z})(z-x)(\bar{z}-x)}\,\delta^{(2)}(\kappa_\parallel)~.
\label{Cphi1}
\end{equation}
Plugging this expression into (\ref{VopenIC}) and performing the algebra on the 
$\Gamma$-matrices in the end we obtain
\begin{equation}
\big\langle \mathcal{V}_{A_1} \big\rangle_{c;I}=(-1)^{I+1}\,2\,\ii\,c\,\kappa_1\,
\delta^{(2)}(\kappa_\parallel)~.
\label{A1C}
\end{equation}
There are no other non-trivial couplings of $c$ since for $A_2$, $\Phi$ and $\Phi_r$ 
the three-point functions vanish at the level of conformal field theory correlators, 
namely
\begin{equation}
\big\langle\mathcal{V}^A(z)\,\mathcal{V}^{\dot{C}}(\bar{z})\,\mathcal{V}_{A_2}(x)
\big\rangle=
\big\langle\mathcal{V}^A(z)\,\mathcal{V}^{\dot{C}}(\bar{z})\,\mathcal{V}_{\Phi}(x)
\big\rangle=
\big\langle\mathcal{V}^A(z)\,\mathcal{V}^{\dot{C}}(\bar{z})\,\mathcal{V}_{\Phi_r}(x)
\big\rangle=0~.
\label{VVC1}
\end{equation}
Obviously this implies that
\begin{equation}
\big\langle \mathcal{V}_{A_2} \big\rangle_{c;I}=
\big\langle \mathcal{V}_{\Phi} \big\rangle_{c;I}=
\big\langle \mathcal{V}_{\Phi_r} \big\rangle_{c;I}=0~.
\end{equation}

\subsubsection*{Correlators with $c'$}
In this case we can be extremely brief since the scalar $c'$ does not couple to any of the massless 
bosonic open string fields. Indeed we have
\begin{equation}
\big\langle \mathcal{V}_{A_1} \big\rangle_{c';I}=
\big\langle \mathcal{V}_{A_2} \big\rangle_{c';I}=
\big\langle \mathcal{V}_{\Phi} \big\rangle_{c';I}=
\big\langle \mathcal{V}_{\Phi_r} \big\rangle_{c';I}=0~.
\end{equation}
The last three equalities clearly follow from (\ref{VVC1}), while the vanishing 
of the coupling of $A_1$ is due to the fact that the $\Gamma$-matrices in 
the numerator of (\ref{Cphi1}) give a zero result when they are contracted with 
the polarization $ (C\,\Gamma_{12})_{A\dot{B}}$. Thus, like $b'$, the scalar $c'$ will also not play any
role in our further analysis.

\section{Continuous parameters of surface operators from world-sheet correlators}
\label{sec:profile}
In this section we provide an interpretation of the non-vanishing couplings
between the closed string massless fields of the twisted sectors and the massless open 
string fields on the fractional D3-branes.

The twisted scalar $b$ of the NS/NS sector
produces a tadpole-like source for the gauge field $A_2$ given in (\ref{VA2fin}), which
depends on the orthogonal momentum to the surface defect. This source, which is localized at the orbifold fixed point where $b$ is defined, gives rise to a non-trivial profile for $A_2$ in the transverse directions: This profile is obtained by computing the Fourier transform 
of the tadpole after including the massless 
propagator
\begin{equation}
\frac{1}{2(|\kappa_\parallel|^2+|\kappa_\perp|^2)}=
\frac{1}{k_1^2+k_2^2+k_3^2+k_4^2}~.
\end{equation}
This procedure is the strict analogue of what has been discussed in 
\cite{DiVecchia:1997vef} for the profile of the gravitational fields emitted by a 
D$p$-brane and in \cite{Billo:2002hm} for the instanton profile of the gauge fields of a 
D3-brane in the presence of D-instantons.

One new feature in this orbifold case is that for functions $f_+$ and $f_-$ which are, respectively, 
even and odd under $\mathbb{Z}_2$, the Fourier transform is given by
\begin{equation}
\begin{aligned}
\mathcal{FT}[f_+](z)&=\int \frac{d^2\kappa_\parallel\,d^2\kappa_\perp}{(2\pi)^2}\,
\cos(\kappa_\perp\!\cdot\!z_\perp)\,
\rme^{\ii\,\kappa_\parallel\cdot
z_\parallel}\,f_+(\kappa)~,\\[1mm]
\mathcal{FT}[f_-](z)&=\int \frac{d^2\kappa_\parallel\,d^2\kappa_\perp}{(2\pi)^2}\,
\ii\,\sin(\kappa_\perp\!\cdot\!z_\perp)\,
\rme^{\ii\,\kappa_\parallel\cdot
z_\parallel}\,f_-(\kappa)~.
\end{aligned}
\label{fourier}
\end{equation}

Let us consider for simplicity a fractional D3-brane type 0. 
Applying the above procedure, the profile of its gauge field $A_2$ in 
configuration space induced by the NS/NS twisted scalar $b$ is
\begin{equation}
\begin{aligned}
A_2&=\int \frac{d^2\kappa_\parallel\,d^2\kappa_\perp}{(2\pi)^2}\,
\ii\,\sin(\kappa_\perp\!\cdot\!z_\perp)
\,\rme^{\ii\,\kappa_
\parallel\cdot
z_\parallel}\,\frac{\big\langle \mathcal{V}_{A_2} \big\rangle_{b;0}}{2(|\kappa_\parallel|^2+|\kappa_\perp|^2)}
\\[1mm]
&=-\ii\,b
\int \frac{d^2\kappa_\perp}{(2\pi)^2}\,\sin(\kappa_\perp\!\cdot\!z_\perp)\,
\frac{\kappa_2}{2|\kappa_\perp|^2}
\end{aligned}
\label{A2prof}
\end{equation}
where in the second line we have used (\ref{VA2fin}) 
with $I=0$ and taken into account the $\delta$-function enforcing 
momentum conservation in the parallel directions to perform the integral 
over $\kappa_\parallel$. This shows that, as anticipated, the propagation of the source is only
in the transverse directions. With a simple calculation we can see that
\begin{equation}
\int \frac{d^2\kappa_\perp}{(2\pi)^2}\,\sin(\kappa_\perp\!\cdot\!z_\perp)\,
\frac{\kappa_2}{2|\kappa_\perp|^2}=\frac{1}{4\pi \bar{z}_2}~,
\end{equation}
so that
\begin{equation}
A_2=-\frac{\ii\,b}{4\pi \bar{z}_2}~.
\label{A2final}
\end{equation}
The component $\bar{A}_2$ of the gauge field also has a non-trivial profile 
which is given by the complex conjugate of (\ref{A2final}). 

As we have seen in the previous section, there are no other tadpole-like sources for 
$A_2$, so that (\ref{A2final}) is the full result. One might think that the R/R scalar 
$c$ can act as a source for the longitudinal component $A_1$ of the gauge 
field in view of (\ref{A1C}). However, if one takes into account the $\delta$-function that 
enforces momentum conservation along the first complex direction, one easily 
realizes that this actually vanishes. Therefore, the vector field is only sourced by the 
NS/NS twisted scalar $b$ which yields (\ref{A2final}) and its complex conjugate. 

In conclusion, the gauge field on a fractional D3-brane of type 0 in 
the $\mathbb{Z}_2$ orbifold acquires the following profile
\begin{equation}
\mathbf{A}=A\cdot dx =A_2\,d\bar{z}_2+\bar{A}_2\,dz_2=-\frac{\ii\,b}{4\pi}\,\left(
\frac{d\bar{z}_2}{\bar{z}_2}-\frac{dz_2}{z_2}\right)=-\frac{b}{2\pi}\,d\theta
\label{profileA2}
\end{equation}
where $\theta$ is the polar angle in the $\mathbb{C}_{(2)}$ plane transverse to the 
defect. If we take a fractional D3-brane of type 1, we obtain the same profile but with an 
overall minus sign due to the different sign in twisted component of the boundary state 
and in the reflection rules (see (\ref{reflexNS})).

Let us now consider the scalar field $\Phi$, the only other open string field that has a 
non-vanishing tadpole produced by $b_+$. Applying the same procedure discussed 
above and using (\ref{VPhiplus}), for a fractional D3-brane of type 0 we obtain
\begin{equation}
\begin{aligned}
\Phi&=\int \frac{d^2\kappa_\parallel\,d^2\kappa_\perp}{(2\pi)^2}\,
\rme^{\ii\,\kappa_\parallel\cdot z_\parallel}\,\ii\,\sin(\kappa_\perp\!\cdot\!z_\perp)\,
\frac{
\big\langle \mathcal{V}_{\Phi} \big\rangle_{b_+;0}}{2(|\kappa_\parallel|^2+|\kappa_\perp|^2)}
\\[1mm]&=b_+
\int \frac{d^2\kappa_\perp}{(2\pi)^2}\,\sin(\kappa_\perp\!\cdot\!z_\perp)\,
\frac{\bar{\kappa}_2}{2|\kappa_\perp|^2}
\,=\,\frac{b_+}{4\pi z_2}~.
\end{aligned}
\end{equation}
Of course, for a fractional D3-brane of type 1 we get the same result with an overall 
minus sign.

It is quite straightforward to generalize these findings to the case of a system made 
of $n_0$ fractional D3-branes of type 0 and $n_1$ fractional D3-branes of type 1, which
describes a gauge theory with group U($n_0+n_1$) broken to the Levi
group U($n_0$)$\times$U($n_1$). In fact, we simply obtain
\begin{subequations}
\begin{align}
\mathbf{A}&=-\frac{b}{2\pi}\,\begin{pmatrix}
\mathbb{1}_{n_0}&0\\
0&-\mathbb{1}_{n_1}
\end{pmatrix}\,d\theta~,\label{An0n1}\\[1mm]
\mathbf{\Phi}&=\frac{b_+}{4\pi}\,\begin{pmatrix}
\mathbb{1}_{n_0}&0\\
0&-\mathbb{1}_{n_1}
\end{pmatrix}\,\frac{1}{z_2}~.
\label{Phin0n1}
\end{align}
\end{subequations}
This is precisely the expected profile for a monodromy defect of GW type.
Comparing with (\ref{Aprofile}) and (\ref{Phiprofile})
we see that the continuous parameters of the surface 
defect are related to the background fields of the NS/NS twisted sector as follows
\begin{equation}
\alpha_I=(-1)^{I+1}\,\frac{b}{2\pi}~,\quad
\beta_I=(-1)^{I}\,\frac{\mathrm{Re}(b_+)}{2\pi}~,\quad
\gamma_I=(-1)^{I}\,\frac{\mathrm{Im}(b_+)}{2\pi}~.
\label{alphabetagamma}
\end{equation}
Notice that in our realization we have $\sum_I\alpha_I=\sum_I\beta_I=\sum_I\gamma_I=0$. 
This is not a limitation since a generic GW solution can always be brought to 
this form by adding a U(1) term proportional to the identity without changing 
the Levi subgroup $\mathrm{U}(n_0)\times\mathrm{U}(n_1)$.

To obtain the profile in the case of special unitary groups we have to remove the overall 
U(1) factor. This is simply done as follows
\begin{subequations}
\begin{align}
\mathbf{A}&~\longmapsto ~\mathbf{A} -\frac{1}{n_0+n_1}\,\big(\Tr \mathbf{A}\big)
\,\mathbb{1}_{n_0+n_1}
\,=\,-\frac{b}{2\pi}\,\begin{pmatrix}
\frac{n_1}{n_0+n_1}\,\mathbb{1}_{n_0}&0\\
0&-\frac{n_0}{n_0+n_1}\,\mathbb{1}_{n_1}
\end{pmatrix}\,d\theta~,\label{An0n1S}\\[1mm]
\mathbf{\Phi}&~\longmapsto ~ \mathbf{\Phi} 
-\frac{1}{n_0+n_1}\,\big(\Tr \mathbf{\Phi}\big)\,\mathbb{1}_{n_0+n_1}
\,=\,\frac{b_+}{4\pi}\,\begin{pmatrix}
\frac{n_1}{n_0+n_1}\,\mathbb{1}_{n_0}&0\\
0&-\frac{n_0}{n_0+n_1}\,\mathbb{1}_{n_1}
\end{pmatrix}\,\frac{1}{z^2}~.
\label{Phin0n1S}
\end{align}
\end{subequations}

Let us now comment on the meaning of the result (\ref{A1C}), which indicates a coupling between the longitudinal component of the gauge field $A_1$ and the twisted scalar $c$ in the R/R sector. This cannot be interpreted 
as a source for the gauge field $A_1$ because it is 
not proportional to the transverse momentum but to the longitudinal one,
which is set to zero by the momentum conserving $\delta$-function. 
However, a different and interesting interpretation is possible. If we multiply
the amplitude (\ref{A1C}) and its complex conjugate by the corresponding polarizations 
of the gauge field, namely $\bar{A}_1$ and $A_1$, the resulting sum 
can be interpreted as an effective interaction term involving the gauge field strength in the 
longitudinal directions. Such a term can be non-zero even in the presence of the
momentum conserving $\delta$-function provided the field strength is kept
fixed. To make this explicit, let us consider a D3-brane of type 0 and
use (\ref{A1C}) for $I=0$. Then we have
\begin{equation}
\bar{A}_1\,\big\langle \mathcal{V}_{A_1} \big\rangle_{c;0} + 
A_1\,\big\langle \mathcal{V}_{\bar{A}_1} \big\rangle_{c;0} = -2\,\ii\,c\,
(\kappa_1\,\bar{A}_1-\bar{\kappa}_1\,A_1)\,\delta^{(2)}(k_\parallel)
=2\,\ii\,c\,\widetilde{F}_0\,\delta^{(2)}(\kappa_\parallel)
\label{amplitude}
\end{equation}
where $\widetilde{F}_0=\bar{\kappa}_1\,A_1-\kappa_1\,\bar{A}_1$ is the (momentum space) 
field strength in the 2$d$ space where the surface defect is extended. 
The Fourier transform of (\ref{amplitude}), computed according to (\ref{fourier}), is
\begin{equation}
\ii\,c\int d^2k_\parallel\,\widetilde{F}_0\,\delta^{(2)}(\kappa_\parallel)\,\times\,2\,\delta^{(2)}(z_\perp)~=~
\frac{\ii\,c}{2\pi}\int d^2x_\parallel\,F_0\,\times\,2\,\delta^{(2)}(z_\perp)
\label{Seff}
\end{equation}
where $F_0$ is the field strength in configuration space. If we assume that this 2$d$ space 
instead of being simply $\mathbb{C}$ is a manifold $D$ where the gauge field strength 
has a non-vanishing first Chern class, then (\ref{Seff}) can be interpreted as
an effective interaction term localized\,%
\footnote{Notice that the term that localizes on the defect placed at the origin is 
$2\,\delta^{(2)}(z_\perp)$, where the factor of 2 compensates the fact that the orbifold 
halves the volume of the transverse space.} on $D$,
meaning that in the path-integral of the underlying (abelian) gauge theory
one has the following phase factor 
\begin{equation}
\exp\left(\frac{\ii\,c}{2\pi}\int_D F_0\right)~.
\label{phase}
\end{equation}
If we extend this argument to a system made of $n_0$ fractional D3-branes
of type 0 and $n_1$ fractional D3-branes of type 1, the phase factor becomes
\begin{equation}
\exp\left(\ii\,\sum_{I}(-1)^{I}\,\frac{c}{2\pi}\int_D \Tr_{\mathrm{U}(n_I)} 
F_I\right)
\label{phasenI}
\end{equation}
which has exactly the same form of the one of the GW monodromy defect
given in (\ref{thetaterm}) with
\begin{equation}
\eta_I =(-1)^{I}\,\frac{c}{2\pi}~.
\label{etaIis}
\end{equation}
In the case of special unitary groups, we have to remove the overall U(1) factor
and this leads to
\begin{equation}
\exp\left(\frac{\ii\,c}{2\pi}\,\frac{n_1}{n_0+n_1}
\int_D \Tr_{\mathrm{U}(n_0)}  F_0 
-\frac{\ii\,c}{2\pi}\,\frac{n_0}{n_0+n_1}\int_D  \Tr_{\mathrm{U}(n_1)}  F_1\right)~.
\label{phaseSUn}
\end{equation}

\section{Conclusions}
\label{secn:concl}

We have shown that a system of $n_0$ fractional D3-branes
of type 0 and $n_1$ fractional D3-branes of type 1 that partially
extend along a $\mathbb{Z}_2$ orbifold, supports a gauge theory 
with a surface defect of the GW type, whose discrete data $(n_0,n_1)$ are encoded 
in the representation of the orbifold group assigned to the fractional D3-branes and 
whose continuous data are encoded in the expectation values of the closed string fields in 
the orbifold twisted sectors according to
\begin{equation}
\left\{\alpha_I,\beta_I,\gamma_I,\eta_I\right\}
= \Big\{(-1)^{I+1}\,\frac{b}{2\pi},(-1)^{I}\,\frac{\mathrm{Re}(b_+)}{2\pi},
(-1)^{I}\,\frac{\mathrm{Im}(b_+)}{2\pi},(-1)^{I}\,\frac{c}{2\pi}\Big\}~.
\label{alphabetagammaeta}
\end{equation}
In the case of special unitary gauge groups, the parameters with $I=0$ must be 
multiplied by $\frac{n_1}{n_0+n_1}$ and those with $I=1$ by $\frac{n_0}{n_0+n_1}$
in order to enforce the decoupling of the overall U(1) factor.

This explicit realization of the continuous parameters of the surface defect in terms
of closed string fields allows us to also discuss how they behave under duality
transformations. To do so we first recall that, from a geometric point of view,
the twisted scalars $b$ and $c$ arise by wrapping the NS/NS and R/R 2-form fields
$B_{(2)}$ and $C_{(2)}$ of Type II B string theory around 
the exceptional 2-cycle $\omega_2$ at the orbifold fixed point
\cite{Bertolini:2000dk,Polchinski:2000mx,Billo:2001vg},
namely
\begin{equation}
b =\int_{\omega_2} B_{(2)}~,\qquad
c =\int_{\omega_2} C_{(2)}~.
\label{bc}
\end{equation}
Using this fact, we can then rewrite the parameters $\alpha_I$ and $\eta_I$ given in
(\ref{alphabetagammaeta}) in the following suggestive way
\begin{equation}
\alpha_I = \frac{(-1)^{I+1} }{2\pi}\int_{\omega_2} B_{(2)}
~,\qquad
\eta_I = \frac{(-1)^I}{2\pi}\int_{\omega_2} C_{(2)}~.
\end{equation}
These formulas, including the relative minus sign, 
are reminiscent of those obtained in \cite{Gomis:2007fi,Drukker:2008wr} 
where a holographic representation of the GW surface defects has been proposed in 
terms of bubbling geometries, which are particular solutions of Type II B supergravity 
with an $AdS_5\times S_5$ asymptotic limit. Our explicit realization in terms of perturbative string 
theory, however, is very different, although the identification of the parameters $\alpha_I$ and $\eta_I$ with the holonomies of the two 2-forms of Type II B is similar.  

The exceptional 2-cycle $\omega_2$ has a vanishing size in the orbifold limit but when 
the orbifold singularity is resolved in a smooth space, it acquires a finite size. The other 
three fields of the twisted NS/NS sector, $b'$ and $b_\pm$, correspond precisely to the 
blow-ups of the orbifold fixed point \cite{Anselmi:1993sm,Polchinski:2000mx}. 
In particular $b'$ is the K\"ahler modulus while $b_\pm$ are the complex structure 
moduli of the blown-up 2-cycle. Hence they are directly 
related to the string-frame metric $G_{\mu\nu}$ of the Type II B string theory.

This geometric interpretation fixes the duality transformations of the twisted
fields since they are inherited from those of the parent Type II B fields from which 
they descend. It is well-known\,%
\footnote{See, for instance, \cite{Becker:2007zj}.}
that under a duality transformation 
$\Lambda=
\bigl(\begin{smallmatrix}
m & n\\
p & q\\
\end{smallmatrix}\bigr) \in \text{SL}(2,\mathbb{Z})$
the two 2-forms rotate among themselves according to
\begin{equation}
\begin{pmatrix}C_{(2)} \\ B_{(2)}\end{pmatrix} ~\longrightarrow ~
\begin{pmatrix}
m & n \\
p & q
\end{pmatrix} 
\begin{pmatrix}C_{(2)} \\ B_{(2)}\end{pmatrix}
\label{BCdual}
\end{equation}
while the string-frame metric $G_{\mu\nu}$ transforms as
\begin{equation}
G_{\mu\nu}~\longrightarrow ~
|p\,\tau+q|\,G_{\mu\nu}
\label{Gdual}
\end{equation}
where $\tau$ is the axio-dilaton field. Therefore, under a duality
$b$ and $c$ rotate as in (\ref{BCdual}) and $b_\pm$ transform as the
metric in (\ref{Gdual}). From this and the identification (\ref{alphabetagammaeta}),
it follows with straightforward manipulations that the surface operator parameters
transform as 
\begin{equation}
\begin{aligned}
(\alpha_I,\eta_I)&~\longrightarrow ~ (q\,\alpha_I-p\,\eta_I,
-n\,\alpha_I+m\,\eta_I)~,\\
(\beta_I,\gamma_I)&~\longrightarrow ~|p\,\tau+q|\,(\beta_I,\gamma_I)
\end{aligned}
\label{Sdualityonabge1}
\end{equation}
Comparing with (\ref{Sdualityonabge}), we see that this is precisely the expected 
behavior of the parameters of the GW defect as originally shown in \cite{Gukov:2006jk}. 
This agreement is an important check of our proposal for the realization of surface 
operators using perturbative string theory.

\vskip 1cm
\noindent {\large {\bf Acknowledgments}}
\vskip 0.2cm
We would like to thank Abhijit Gadde, Dileep Jatkar, Naveen Prabhakar, Madhusudhan Raman and Ashoke Sen for helpful discussions and correspondence and Renjan John for collaboration at an early stage of the project. The work of A.L. is partially supported by ``Fondi Ricerca Locale dell'Universit\`a  del Piemonte Orientale".
\vskip 1cm
\begin{appendix}
\section{Dirac matrices}
\label{spinorconventions}

In this appendix, we define in detail our conventions for the Dirac matrices used in the main text.

\subsection{4$d$}
\label{twistedNSspinors}
We consider the 4$d$ Euclidean space spanned by the coordinates $x_m$ with
$m\in \{3,4,5,6\}$. These are the real coordinates corresponding to the complex coordinates
$z_2$ and $z_3$ (see (\ref{coords})) along which the $\mathbb{Z}_2$ orbifold acts.

An explicit realization of the Dirac matrices $\gamma_m$ satisfying the 
4$d$ Euclidean Clifford algebra
\begin{equation}
\{\gamma_{m}, \gamma_{n}\} = 2\delta_{mn}~,
\end{equation}
is given by
\begin{align}
\gamma_3=\begin{pmatrix}
0 & \tau_1\\
\tau_1& 0\\
\end{pmatrix}~,\quad
\gamma_4=\begin{pmatrix}
0 & -\tau_2\\
-\tau_2& 0\\
\end{pmatrix}~,\quad
\gamma_5=\begin{pmatrix}
0 & \tau_3\\
\tau_3& 0\\
\end{pmatrix}~,\quad
\gamma_6=\begin{pmatrix}
0 & \ii\,\mathbb{1}_2\\
-\ii\,\mathbb{1}_2& 0\\
\end{pmatrix}
\end{align}
where $\tau_c$ are the usual Pauli matrices and $\mathbb{1}_2$ is the $2\times 2$ identity matrix.

The chirality matrix $\widehat{\gamma}$ is given by
\begin{equation}
\widehat{\gamma}=-\gamma_3\gamma_4\gamma_5\gamma_6=\begin{pmatrix}
\mathbb{1}_2 & 0\\
0& -\mathbb{1}_2\\
\end{pmatrix}
~.
\label{4dchirality}
\end{equation}
This shows that in this basis a 4$d$ Dirac spinor is written as
\begin{equation}
\begin{pmatrix}
S^\alpha\\
S^{\dot{\alpha}}
\end{pmatrix}
\end{equation}
where $\alpha$ and $\dot{\alpha}$ label, respectively, the chiral and anti-chiral components.

Finally, the charge conjugation matrix $\widehat{C}$ is given by
\begin{equation}
\widehat{C}=\begin{pmatrix}
\epsilon & 0\\
0&-\epsilon
\end{pmatrix}
\label{chargeconj4}
\end{equation}
where $\epsilon=-\ii\,\tau_2$ (see (\ref{epsilon})), and is such that
\begin{equation}
\widehat{C}\,\gamma_m\,\widehat{C}^{-1}= (\gamma_m)^{\mathtt{t}}
\end{equation}
where $\mathtt{t}$ denotes the transpose.

\subsection{6$d$}
\label{twistedRspinors}
We consider the 6$d$ Euclidean space spanned by the coordinates $x_M$ with
$M\in \{1,2,7,8,9,10\}$. These are the real coordinates corresponding to the complex coordinates
$z_1$, $z_4$ and $z_5$ (see (\ref{coords})) that are transverse to the $\mathbb{Z}_2$ orbifold.

An explicit realization of the Dirac matrices $\Gamma_M$ satisfying the 
6$d$ Euclidean Clifford algebra
\begin{equation}
\{\Gamma_{M}, \Gamma_{N}\} = 2\delta_{MN}~,
\end{equation}
is given by
\begin{equation}
\begin{aligned}
\Gamma_1&=\begin{pmatrix}
0 & 0&-\ii\, \mathbb{1}_2 & 0\\
0 & 0& 0 & -\ii\,\mathbb{1}_2\\
\ii\,\mathbb{1}_2 & 0& 0 & 0\\
0 & \ii\,\mathbb{1}_2& 0 & 0\\
\end{pmatrix}~,\quad
\quad\quad
\Gamma_2=\begin{pmatrix}
0 & 0& \tau_3 & 0\\
0 & 0& 0 & -\tau_3\\
\tau_3 & 0& 0 & 0\\
0 & -\tau_3& 0 & 0\\
\end{pmatrix}~,\\[2mm]
\Gamma_7&=\begin{pmatrix}
0 & 0& -\tau_2 & 0\\
0 & 0& 0 & \tau_2\\
-\tau_2 & 0& 0 & 0\\
0 & \tau_2& 0 & 0\\
\end{pmatrix}~,\quad
\qquad\quad~~~
\Gamma_8=\begin{pmatrix}
0 & 0& \tau_1 & 0\\
0 & 0& 0 & -\tau_1\\
\tau_1 & 0& 0 & 0\\
0 & -\tau_1& 0 & 0\\
\end{pmatrix}~,\\[2mm]
\Gamma_9&=\begin{pmatrix}
0 & 0& 0 & -\ii\,\mathbb{1}_2\\
0 & 0& \ii\,\mathbb{1}_2 & 0\\
0 & -\ii\,\mathbb{1}_2& 0 & 0\\
\ii\,\mathbb{1}_2 & 0& 0 & 0\\
\end{pmatrix}~,\quad
\quad~~~
\Gamma_{10}=\begin{pmatrix}
0 & 0& 0 & \mathbb{1}_2\\
0 & 0& \mathbb{1}_2 & 0\\
0 & \mathbb{1}_2& 0 & 0\\
\mathbb{1}_2 & 0& 0 & 0\\
\end{pmatrix}~.\\
\end{aligned}
\end{equation}
The chirality matrix $\widehat{\Gamma}$ is 
\begin{equation}
\widehat{\Gamma}=\ii\,\Gamma_1\Gamma_2\Gamma_7\Gamma_8\Gamma_9\Gamma_{10}
=\begin{pmatrix}
\mathbb{1}_2 & 0 & 0 & 0\\
0& \mathbb{1}_2& 0 &0\\
0& 0&-\mathbb{1}_2& 0\\
0& 0& 0& -\mathbb{1}_2
\end{pmatrix}
~.
\label{6dchirality}
\end{equation}
This shows that in this basis a 6$d$ Dirac spinor is written as
\begin{equation}
\begin{pmatrix}
S^A\\
S^{\dot{A}}
\end{pmatrix}
\end{equation}
where $A$ and $\dot{A}$ label, respectively, the chiral and anti-chiral components.

The charge conjugation matrix $C$ is
\begin{equation}
C=\begin{pmatrix}
0&0&0&\epsilon\\
0&0&\epsilon&0\\
0&-\epsilon&0&0\\
-\epsilon&0&0&0
\end{pmatrix}
\label{chargeconjugation6d}
\end{equation}
where, as before, $\epsilon=-\ii\,\tau_2$. The above charge conjugation matrix is such that
\begin{equation}
C\,\Gamma_M\,C^{-1}=- (\Gamma_M)^{\mathtt{t}}~.
\end{equation}

\section{Reflection rules}
\label{app:gluing}
When the world-sheet of the closed string has a boundary, there are 
non trivial 2-point functions between the left and right moving
parts. We are interested in computing these 2-point functions for the massless
fields of the twisted sectors when the boundary is
created by a fractional D3-brane of type $I$ in the $\mathbb{Z}_2$ orbifold discussed in 
Section~\ref{sec:fD3}. 

In the boundary state formalism (see for instance the reviews \cite{DiVecchia:1999mal,DiVecchia:1999fje}) the boundary created by a D-brane is the unit circle, {\it{i.e.}} the set of points corresponding to the world-sheet time $\tau=0$ where the boundary state is inserted. The points inside the unit circle define the disk $\mathbb{D}$. When we insert a closed string inside 
$\mathbb{D}$, the left and right moving modes are reflected at the boundary, and a 
non-vanishing correlator between them arises. For example, considering the twisted NS/NS 
sector and in particular the massless states described by the vertex operators (\ref{NSvertexops})
in the presence of a fractional D3-brane of type $I$, we have
\begin{equation}
\begin{aligned}
\langle \mathcal{V}^\alpha(w)\,\widetilde{\mathcal{V}}^\beta(\bar{w})\rangle_{I}&=
\langle T;I|\,\mathcal{V}^\alpha(w)\,\widetilde{\mathcal{V}}^\beta
(\bar{w})\,|0\rangle|\widetilde{0}\rangle\\[1mm]
&=
(-1)^{I}\,\,{}_{\mathrm{NS}}\langle T|\,\mathcal{V}^\alpha(w)\,\widetilde{\mathcal{V}}^\beta
(\bar{w})\,|0\rangle|\widetilde{0}\rangle
\end{aligned}
\label{2point V1}
\end{equation}
for $w$ and $\bar{w}\in \mathbb{D}$. Here we have used the boundary state 
to represent the fractional D3-brane of type $I$ (see (\ref{fracD3s}))
and taken into account that only the NS component of its twisted part is relevant for the calculation. As in the main text, 
$|0\rangle$ and $\widetilde{|0\rangle}$ denote the left and right vacua.

On the other hand, conformal invariance implies that the disk 2-point function 
of $\cV^\alpha$ and $\widetilde{\cV}^\beta$, which are conformal fields of weight 1, has the
following form
\begin{equation}
\langle \mathcal{V}^\alpha(w)\,\widetilde{\mathcal{V}}^\beta(\bar{w})\rangle_{I}=
\frac{M^{\alpha\beta}_I}{(1-w\bar{w})^{2}}
\label{2point V2}
\end{equation}
where $M^{\alpha\beta}_I$ is a constant to be determined.
Combining (\ref{2point V1}) and (\ref{2point V2}), we easily see that
\begin{equation}
M^{\alpha\beta}_I=\lim_{w\to 0}\,\lim_{\bar{w}\to 0} \,
\langle \mathcal{V}^\alpha(w)\,\widetilde{\mathcal{V}}^\beta(\bar{w})\rangle_{I}
=(-1)^{I}\,\,{}_{\mathrm{NS}}\langle T|\alpha\rangle |\widetilde{\beta}\rangle
\label{Malphabetais}
\end{equation}
where $|\alpha\rangle$ and $|\widetilde{\beta}\rangle$ are the left and right ground
states defined in (\ref{bdryTNS1}). Thus, the disk 2-point function (\ref{2point V2}) becomes
\begin{equation}
\langle \mathcal{V}^\alpha(w)\,\widetilde{\mathcal{V}}^\beta(\bar{w})\rangle_{I}
=(-1)^{I}\,\frac{{}_{\mathrm{NS}}\langle T|\alpha\rangle
|\widetilde{\beta}\rangle}{(1-w\bar{w})^2}~.
\end{equation}
Let us now map this result to the complex plane by means of the Cayley map
\begin{equation}
w=\frac{z-\ii}{z+\ii}~,\quad
\bar{w}=\frac{\bar{z}+\ii}{\bar{z}-\ii}~.
\end{equation}
Notice that $w$ is mapped to the upper half-complex plane and $\bar{w}$ to the lower half.
Then, we have
\begin{equation}
\begin{aligned}
\langle \mathcal{V}^\alpha(z)\,\widetilde{\mathcal{V}}^\beta(\bar{z})\rangle_{I}
&=\langle \mathcal{V}^\alpha(w)\,\widetilde{\mathcal{V}}^\beta(\bar{w})\rangle_{I}
\,\frac{dw}{dz} \,\frac{d\bar{w}}{d\bar{z}}
=(-1)^{I+1}\,\frac{{}_{\mathrm{NS}}\langle
T|\alpha\rangle|\widetilde{\beta}\rangle}{(z-\bar{z})^2}~.
\end{aligned}
\end{equation}
Comparing with (\ref{VVtilde}) and using the so-called doubling trick, we are led to 
introduce the following reflection rule 
\begin{equation}
\widetilde{\mathcal{V}}^\beta(\bar{z})\longrightarrow
(R_I)^\beta_{~\gamma}\,\mathcal{V}^\gamma(\bar{z})
\end{equation}
with
\begin{equation}
(R_I)^\beta_{~\gamma}=(-1)^{I}\,\widehat{C}_{\gamma\alpha}
\,{}_{\mathrm{NS}}\langle T|\alpha\rangle |\widetilde{\beta}\rangle
=(-1)^{I+1}\,\varepsilon_{\gamma\alpha}
\,{}_{\mathrm{NS}}\langle T|\alpha\rangle |\widetilde{\beta}\rangle
\label{reflexapp}
\end{equation}
where in the second step we have used the fact that the chiral part of the charge conjugation matrix 
is $\epsilon$ (see \ref{chargeconj4})).
Using the expression (\ref{bdryTNS}) for the twisted boundary state in the NS/NS sector, it
is easy to show that
\begin{equation}
{}_{\mathrm{NS}}\langle T|\alpha\rangle |\widetilde{\beta}\rangle=(\gamma_4\gamma_3\widehat{C}^{-1})^{\beta\alpha}~.
\label{Talphabetaapp}
\end{equation}
Inserting this into (\ref{reflexapp}), we find
\begin{equation}
(R_I)^\beta_{~\gamma}=(-1)^I
(\gamma_4\gamma_3)^\beta_{~\gamma}
\label{reflexNSapp}
\end{equation}
in agreement with (\ref{reflexNS}) of the main text.

The reflection matrix for the R sector can be obtained in the same way. Indeed, in the presence of
a D3-brane of type $I$ the left and right-moving vertex operators $\cV^A$ and
$\widetilde{\cV}^{\dot{B}}$ have the following 2-point function 
\begin{equation}
\langle \mathcal{V}^A(z)\,\widetilde{\mathcal{V}}^{\dot{B}}(\bar{z})\rangle_{I}
=(-1)^{I+1}\,\frac{{}_{\mathrm{R}}\langle
T|A\rangle|\widetilde{\dot{B}}\rangle}{(z-\bar{z})^2}~.
\end{equation}
Comparing with (\ref{VVtildeR}), we are led to introduce the reflection rule
\begin{equation}
\widetilde{\mathcal{V}}^{\dot{B}}(\bar{z})\longrightarrow
(R_I)^{\dot{B}}_{~\dot{C}}\,\mathcal{V}^{\dot{C}}(\bar{z})
\end{equation}
such that
\begin{equation}
(R_I)^{\dot{B}}_{~\dot{C}}=(-1)^{I+1}\,{C}_{\dot{C}A}
\,{}_{\mathrm{R}}\langle T|A\rangle |\widetilde{\dot{B}}\rangle~.
\label{reflexappR}
\end{equation}
{From} the expression (\ref{bdryTR}) for the twisted boundary state in the R/R sector, one
can show that
\begin{equation}
{}_{\mathrm{R}}\langle T|A\rangle |\widetilde{\dot{B}}\rangle
=(\Gamma_2\Gamma_1{C}^{-1})^{\dot{B}A}~.
\label{Talphabeta}
\end{equation}
Inserting this into (\ref{reflexappR}), we therefore find
\begin{equation}
(R_I)^{\dot{B}}_{~\dot{C}}=(-1)^I
(\Gamma_1\Gamma_2)^{\dot{B}}_{~\dot{C}}
\label{reflexRapp}
\end{equation}
in agreement with (\ref{reflexR}) of the main text.

\end{appendix}
\endgroup

\providecommand{\href}[2]{#2}\begingroup\raggedright\endgroup


\end{document}

%% file: tadpole.pdf_tex
\begingroup%
  \makeatletter%
  \providecommand\color[2][]{%
    \errmessage{(Inkscape) Color is used for the text in Inkscape, but the package 'color.sty' is not loaded}%
    \renewcommand\color[2][]{}%
  }%
  \providecommand\transparent[1]{%
    \errmessage{(Inkscape) Transparency is used (non-zero) for the text in Inkscape, but the package 'transparent.sty' is not loaded}%
    \renewcommand\transparent[1]{}%
  }%
  \providecommand\rotatebox[2]{#2}%
  \newcommand*\fsize{\dimexpr\f@size pt\relax}%
  \newcommand*\lineheight[1]{\fontsize{\fsize}{#1\fsize}\selectfont}%
  \ifx\svgwidth\undefined%
    \setlength{\unitlength}{135bp}%
    \ifx\svgscale\undefined%
      \relax%
    \else%
      \setlength{\unitlength}{\unitlength * \real{\svgscale}}%
    \fi%
  \else%
    \setlength{\unitlength}{\svgwidth}%
  \fi%
  \global\let\svgwidth\undefined%
  \global\let\svgscale\undefined%
  \makeatother%
  \begin{picture}(1,0.59257385)%
    \lineheight{1}%
    \setlength\tabcolsep{0pt}%
    \put(0.82935067,0.31011491){\color[rgb]{0,0,0}\makebox(0,0)[lt]{\lineheight{0}\smash{\begin{tabular}[t]{l}
    \end{tabular}}}}%
    \put(0,0){\includegraphics[width=\unitlength,page=1]{tadpole.pdf}}%
    \put(0.06179471,0.37063218){\color[rgb]{0,0,0}\makebox(0,0)[lt]{\lineheight{1.25}\smash{\begin{tabular}[t]{l}\textbf{$\cV_{\mathrm{open}}$}\end{tabular}}}}%
    \put(0.56513473,0.21771536){\color[rgb]{0,0,0}\makebox(0,0)[lt]{\lineheight{1.25}\smash{\begin{tabular}[t]{l}\textbf{$b\,\cV_b$}\end{tabular}}}}%
    \put(0.73370991,0.52739252){\color[rgb]{0,0,0}\makebox(0,0)[lt]{\lineheight{1.25}\smash{\begin{tabular}[t]{l}\textbf{$\mathrm{D3}_I$}\end{tabular}}}}%
    \put(0,0){\includegraphics[width=\unitlength,page=2]{tadpole.pdf}}%
    \put(0.0464934,0.19462331){\color[rgb]{0,0,0}\makebox(0,0)[lt]{\lineheight{1.25}\smash{\begin{tabular}[t]{l}\textbf{$\vec{k}_\perp$}\end{tabular}}}}%
  \end{picture}%
\endgroup%